\documentclass{aa_arxiv}

\usepackage{graphicx}
\usepackage{txfonts}
\usepackage{mathtools}
\usepackage{amsmath}
\usepackage[utf8]{inputenc}
\def\fd{$\,.\!^{\rm \!d}$}
\usepackage[colorlinks,citecolor=blue]{hyperref}
\hypersetup{
    colorlinks = true,
    linkcolor = blue,
    anchorcolor = blue,
    citecolor = blue,
    filecolor = blue,
    urlcolor = blue
    }

\begin{document}

   \title{YY Hya and its interstellar environment}

   \author{Stefan Kimeswenger\inst{1,2}
   \and John R. Thorstensen\inst{3}
   \and Robert A. Fesen\inst{3}
   \and Marcel Drechsler\inst{4}
   \and     Xavier Strottner\inst{5}
   \and     Maicon Germiniani\inst{6}
   \and    Thomas Steindl\inst{1}
   \and    Norbert Przybilla\inst{1}
   \and Kathryn E. Weil\inst{7}
   \and Justin Rupert\inst{8}
          }

   \institute{
Institut f{\"u}r Astro- und Teilchenphysik, Leopold--Franzens Universit{\"a}t Innsbruck, Technikerstr.~25, 6020 Innsbruck, Austria\protect\newline\email{Stefan.Kimeswenger@uibk.ac.at} %\protect\newline\email{Thomas.Steindl@uibk.ac.at} %\protect\newline\email{Norbert.Przybilla@uibk.ac.at}
\and
Instituto de Astronom{\'i}a, Universidad Cat{\'o}lica del Norte, Av.~Angamos 0610, Antofagasta, Chile
\and
Department of Physics and Astronomy, 6127 Wilder Laboratory, Dartmouth College, Hanover, NH 03755-3528, USA
%\protect\newline\email{John.R.Thorstensen@dartmouth.edu}
%\protect\newline\email{Robert.A.Fesen@dartmouth.edu}
\and
Sternwarte B{\"a}renstein, Feldstraße 17, 09471 B{\"a}renstein, Germany%\protect\newline\email{epost@marcel-drechsler.de}
\and
Montfraze, 01370 Saint Etienne Du Bois
France%\protect\newline\email{leonberx@hotmail.fr}
\and
Leopoldo Stadtlober Street 49, 89871-000 Serra Alta, Brazil
%\protect\newline\email{germinianimaicon@icloud.com}
\and
Department of Physics and Astronomy, Purdue University, 525 Northwestern Avenue, West Lafayette, IN 47907 USA
%\protect\newline\email{keweil@purdue.edu}
\and
MDM Observatory, Kitt Peak National Observatory, 950 N. Cherry Ave., Tucson, AZ 85719, USA
%\protect\newline\email{jurupert@umich.edu}
    }

   \date{Accepted: October 8$^{\rm th}$, 2021}

  \abstract
  % context heading (optional)
   {During a search for previously unknown Galactic emission nebulae, we discovered a faint $36\arcmin$ diameter H$\alpha$ emission nebula centered around the periodic variable YY~Hya. Although this star has been classified as RR-Lyr variable, such a stellar classification is inconsistent with the formation of such a large nebula especially if at YY~Hya's estimated Gaia distance of $\simeq$450 pc. GALEX image data also shows YY~Hya to have a strong UV excess, suggesting the existence of a hot, compact binary companion.}
  % aims heading (mandatory)
  {We aim to clarify the nature of YY~Hya and its nebula.}
  % {We aim here to determine a more likely model of the nature of YY~Hya and its nebula.}
  % methods heading (mandatory)
   {In addition to our discovery image data, we obtained a  $2.\!\!^\circ5 \times 2.\!\!^\circ5$ image mosaic of the whole region with CHILESCOPE facilities and time-series spectroscopy at MDM observatory.  Also, we used data from various space missions to derive an exact orbital period and a spectral energy distribution (SED). The binary star model code Binary Maker 3 (BM3), and Kurucz ATLAS9 stellar atmospheres were used to derive a synthetic light curve and a model SED of the compact binary system, respectively.}
  % results heading (mandatory)
   {We find that YY~Hya is a compact binary system containing a K dwarf star which is strongly irradiated by a hot white dwarf {(WD)} companion. The spectral characteristics of the emission lines, visible only during maximum {light of the perfectly sinusoidal optical} light {curve, when the side of the K star fully illuminated by the WD points to the observer}, shows signatures much like members of the BE UMa variable family. These are post common envelope pre-cataclysmic variables. However the companion star here is more massive than found in other group members and the progenitor of the white dwarf must have been a 3 to 4 $M_\odot$ star. The nebula seems to be an ejected common envelope shell with a mass in the order of one $M_\odot$ and an age of 500\,000 years. This makes it to be the biggest hitherto known such shell. Alignment of neighboring nebulosities some $45\arcmin$ to the northeast and southwest of YY~Hya suggests that the system had strong bipolar outflows.
   We briefly speculate it might be related to the 1065 BP ``guest-star'' reported in ancient Chinese records as well.}{}
  % conclusions heaiding (optional), leave it empty if necessary
   %{%conclusions are optional
   %conclusions}

%https://www.aanda.org/for-authors/latex-issues/information-files#pop
\sloppy
   \maketitle
%
%-------------------------------------------------------------------

\section{Introduction}
\label{sec:intro}

\begin{figure*}[t!]
%% option sidecaption
%% \sidecaption
%% {
%% \includegraphics[width=59.9mm]{HA_image_BW_marked_new.png}\phantom{X}\includegraphics[width=59.9mm]{index4.png}
%% }
%% option full page
\centerline{
\includegraphics[width=88.9mm]{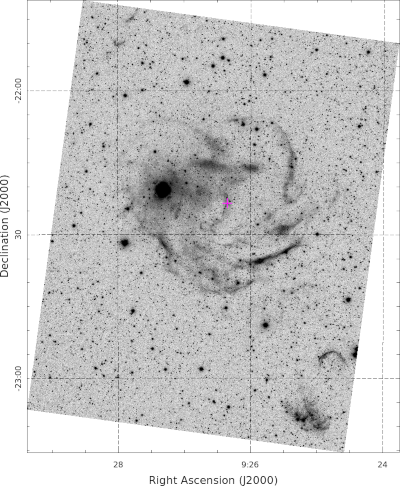}\phantom{X}\includegraphics[width=88.9mm]{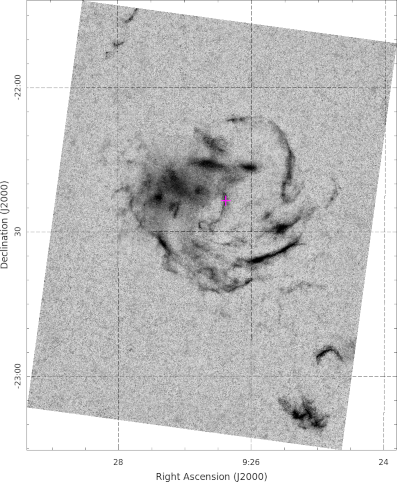}
}
\caption{The H$\alpha$ image taken at Serra Alta, Brazil of the region around YY~Hya. (Left:) The original image; (right) the image after star removal procedure (see text). The inner main nebula has a diameter of 36\arcmin. Clearly visible are in the northeast and in the southwest corner exactly vis-{\`a}-vis emission structures both 47\arcmin~ from the variable star. The northeast structure is just at the image edge and partly cut off. }
\label{fig:nebula}
\end{figure*}

In a systematic search for previously unrecognized Galactic nebulosities on sky survey plates downloaded from the \mbox{SuperCOSMOS} facility archive in Edinburgh\footnote{\url{http://www-wfau.roe.ac.uk/sss/pixel.html}}  \citep{SuperCOSMOS}, we discovered a small bow-like emission structure
near \mbox{$\alpha$(J2000) = 09:25:50}, \mbox{$\delta$(J2000) = $-22$:23:00}. Follow-up images taken with a small telescope facility at Serra Alta, Brazil showed that this feature is the brightest part of a much larger (36\arcmin diameter), highly structured and nearly circular H$\alpha$ nebula centered on the variable star \mbox{\object{YY Hya}}  (\mbox{$\alpha$ = 09:26:20.596}, \mbox{$\delta$ = $-22$:23:38.38}; \mbox{$l$ = 252\fdg81}, \mbox{$b$ = +19\fdg94}). We also found two additional emission structures, located $\sim$45\arcmin\ northeast and southwest of the center (see Fig.\,\ref{fig:nebula}). First thought to be a possible planetary nebula (PN), it was assigned the identifier StDr Object 20 (PN G252.8+19.9) in the HASH PN data base \citep{HASH2016,HASH2017}.

YY~Hya was discovered nearly 85 years ago as Harvard Variable\,No.\,7525 \citep{discover} and classified as an RR Lyr star, called \emph{Cluster} variable in the literature at that time.  \citet{naming} assigned the name YY~Hya shortly after that.
% York study is irrelevant to this paper.
%\citet{CLOUD_DIST} used its RR Lyr classification to help derive %distances to high velocity clouds in the Galaxy without a further %study of the star itself. Later,
\citet{CATALINA_RR}, using data from the Catalina Real-time Transient Survey \citep[CRTS,][]{CATALINA}, determined a period of 0.33479 d, and classified the star as a c-type RR Lyr (RRc), based on its almost perfectly sinusoidal light curve and high amplitude of nearly one magnitude.
% Likewise, the WISE reference is not needed.
%In the databases of the Wide-field Infrared Survey Explorer %\citep[WISE,][]{WISE}, YY~Hya appears as WISE\ J092620.5-222338 with a %RR Lyr classification  apparently copied from previous literature %resources into the WISE catalog of periodic variable stars %\citep{WISE_VAR}.
The  RRc classification implies an absolute magnitude range from $0\fm85 < M_{\rm V} < 0\fm55$ \citep{RR_MAG1,RR_MAG2}. This led to an initial distance estimate of $4.05 < D_{\rm RR} < 4.75\,{\rm kpc}$.

However, recent parallax measurements indicate a much shorter distance, inconsistent with any kind of RR Lyr star. The inverse of the parallax from Data Release 2 (DR2) of the Global Astrometric Interferometer for Astrophysics \citep[Gaia; ][]{GAIA_DR2} is 443$\pm$6 pc, about 10 times smaller than that expected for an RR Lyr star. The more recent Gaia early Data Release 3  \citep[eDR3;][]{GAIA_eDR3} gives 456$\pm$3\,pc. %for the same quantity.
Although this new value does not overlap within the 1\,$\sigma$ errors with the DR2 result, we will adopt that value as it is based on more extensive data and has correspondingly smaller uncertainties.
However, up to now, the Gaia analysis solves only for the star's position, parallax, and proper motions. Moving due to binary orbits is not yet included in Gaia as a source of possible systematic errors. But in the case of the system here as we see later, the whole orbit extends only to 0.03\,mas. That is about 1\,$\sigma$ of the given parallax error in Gaia eDR3. The distance by Gaia thus seems to be still reliable. However the orbital motion might be the cause for the significant shift between the two major data releases. Thus the realistic error, assuming some systematic component, is more likely in the order of 10\,pc.

%\added{Not only the parallax errors, but also all other quality flags in the GAIA DR2 and eDR3 data base show a very good astrometric solution for this source. Thus further on we use the distance of 456\,pc throughout this investigation.}\\
%No further detailed investigations about the star itself were found in the literature. Moreover no previous information about the nebula were found by us.

In this paper, we present follow-up imaging and time series spectroscopy, along with photometric data collected from various literature and data base resources, as well as an analysis of photometry from the Transiting Exoplanet Survey Satellite \citep[TESS,][]{TESS}.
We have used these data to classify the system using radial velocities, spectral energy distribution (SED) and the UV excess found by the  Galaxy Evolution Explorer \citep[GALEX,][]{GALEX_1}.
We furthermore put the kinematic behavior of the system into the Galactic context and derive detailed parameters for the progenitor stars.
Finally, we discuss a possible link to a medieval ``guest-star'' observation reported by Chinese observers in this direction on the sky \citep{Hsi1957,Ho1962}.

\section{Data and Reductions}
\label{sec:sec2}

\subsection{H$\alpha$ Imaging}

%The attention to the object was drawn due to the discovery of the brightest southwestern arc of the nebula on the digital copy of the ESO-R sky survey plate  \citep{ESOR} from SuperCOSMOS \citep{SuperCOSMOS}.

H$\alpha$ discovery images were taken in January 2020 in Serra Alta, Brazil\footnote{Serra Alta, Brazil: 53\fdg0426 W, 26\fdg7285 S}, using a 115mm F7 TS Proline Triple APO refractor from Teleskop--Service Ransburg\footnote{\url{https://www.teleskop-express.de/}} plus a focal reducer resulting in a final f/5.5  The detector was a ZWO\footnote{\url{https://astronomy-imaging-camera.com/}} Camera ASI 1600 MM-Cool equipped with a 4656 $\times$ 3520, 3.8$\mu$m pixel CMOS Panasonic MN34230 and an OPTOLONG\footnote{\url{https://www.optolong.com/}} 7 nm wide H$\alpha$ filter. This resulted in a pixel scale of 1\farcs24 pixel$^{-1}$ and a field of view (FOV) of 1\fdg6 $\times$ 1\fdg21.

The image shown in Fig.\,\ref{fig:nebula} is a composite of 120 $\times$ 600 s exposures for a total exposure time of 20 hours. A set of matching broadband red filter images were also obtained in order to remove the stellar background. These red filter images were stretched manually by a nonlinear intensity curve to correct for differences in the calibration and were finally subtracted from the H$\alpha$ image.

\begin{figure*}[t!]
%% \sidecaption
%%{
%% \includegraphics[width=120mm,height=120mm]{HA_CHILE_part.png}
%% }
%% \full page option
{
\includegraphics[width=180mm,height=180mm]{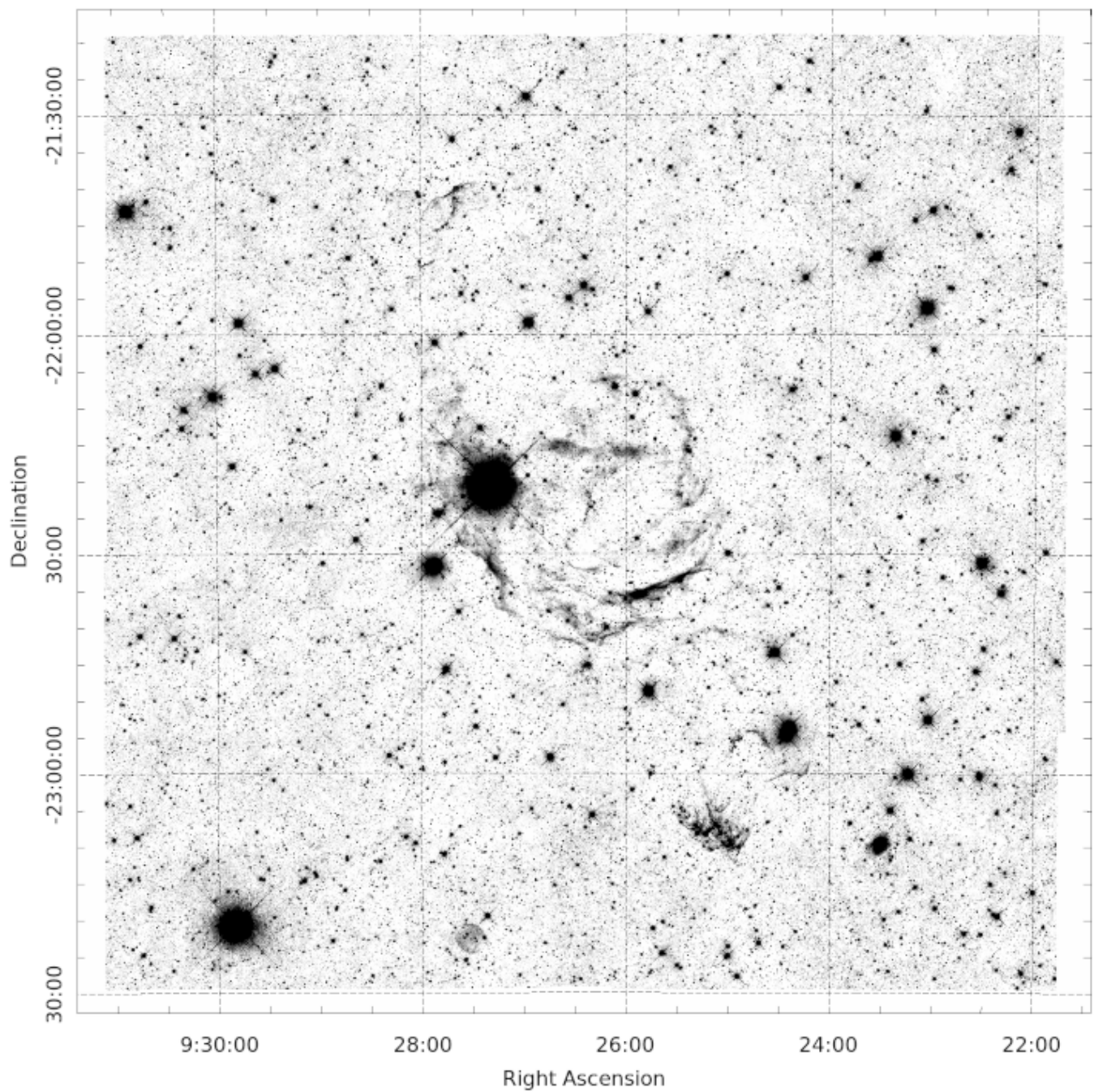}
}
\caption{The H$\alpha$ image mosaic obtained at CHILESCOPE with a linear gray-scale mapping with surface brightness from zero to $1.5\,\,10^{-16}\,\,$erg\,cm$^{-2}$\,s$^{-1}$\,arcsec$^{-1}$. The stellar limiting magnitude is about Gaia RP $\approx$ 21$^{\rm m}\!\!.\,$8.  The round nebula southwards is not related to our target but the beforehand known planetary nebula StDr 47
(PN~G253.7+19.4, 09$^{\rm h}$27$^{\rm m}$31$^{\rm s}\!\!.\,$46, -23\degr22\arcmin34$''\!\!.\,$40). }
\label{fig:chilescope}
\end{figure*}

The resulting images shows a highly structured H$\alpha$ nebula with a diameter of $\sim$36\arcmin. In addition, two smaller nebulosities in opposite directions roughly
 $45\arcmin$ to the northeast and southwest from YY~Hya are also visible. Examination of copies of the original photographic sky surveys ESO-R \citep{ESOR} shows no more evidence for the nebula other than the small southwestern arc already seen on the  H-compressed digital scans. Likewise,  no part of the nebula is visible at the SERC-J  \citep{SERCJ} plate copy, suggesting that the nebula does not have a significant \mbox{[\ion{O}{iii}]}
emission.

To investigate further, we booked time at the CHILESCOPE remotely controlled commercial
observatory facilities.
Details on the location, the instrumentation used and the calibration are given in Appendix\ \ref{app:mosaic}.
%The CHILESCOPE is a remotely controlled commercial
%observatory located in the Chilean Andes.
%We used the two f/3.8 50cm~Newtonian telescopes equipped with
%4\,K\,$\times$\,4\,K FLI PROLINE %16803\footnote{\url{https://www.flicamera.com/}} CCD cameras, yielding
%to a pixel scale of 0\farcs963 pixel$^{-1}$ and a FOV of 1\fdg096 $\times$ %1\fdg096.
Images with  H$\alpha$, [\ion{O}{iii}], and [\ion{S}{ii}] filters from Astrodon\footnote{\url{https://astrodon.com/}} were obtained. However, 30 minutes of exposure in the [\ion{O}{iii}] and [\ion{S}{ii}] bands centered at the main nebula showed no detection of even the brightest filaments. Thus the campaign, lasting from March 9$^{\rm th}$ to June 13$^{\rm th}$ 2021, focused finally completely on the H$\alpha$ imaging.
The aim was the full coverage of the field and the possible detection of further distant structures and a calibrated intensity estimate.
But beside the already mentioned structures, no other distant structures related YY~Hya and its nebula could be identified.
Figure~\ref{fig:chilescope} shows the mosaic of the 294 H$\alpha$ images with 1200 seconds exposure time each (98 hours integration time). The brightest nebular filaments are about $4\,\,10^{-17}\,\,$erg\,cm$^{-2}$\,s$^{-1}$\,arcsec$^{-2}$. The faintest visible structures at 1.5\,$\sigma$ to the background $rms$ of single pixels are around $8\,\,10^{-18}\,\,$erg\,cm$^{-2}$\,s$^{-1}$\,arcsec$^{-2}$.
%The details of the observations and the calibration are given in Appendix~\ref{app:mosaic}.

\begin{figure*}
\begin{center}
\includegraphics[height=64mm]{fig_03_a.png}
\includegraphics[height=64mm]{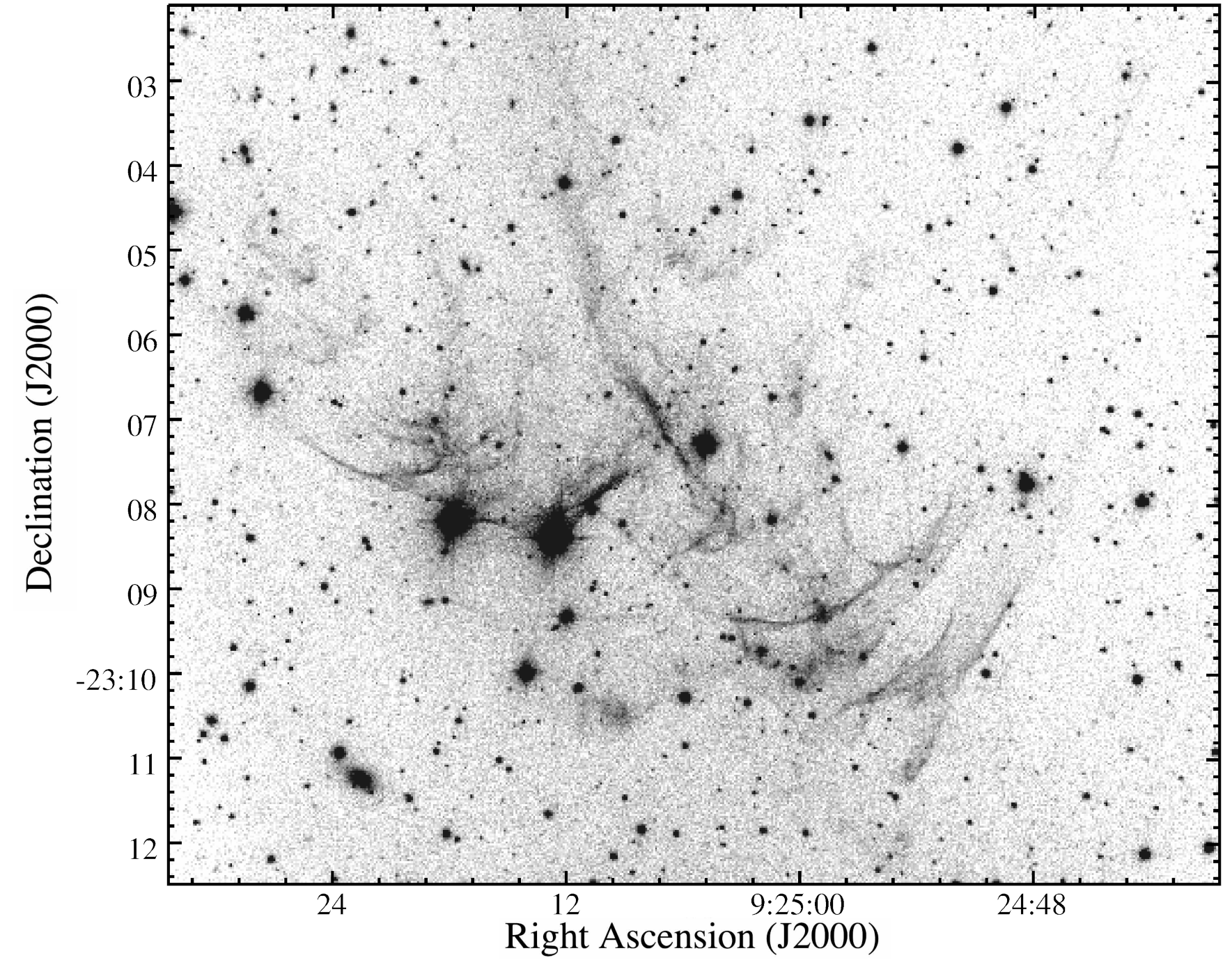}
\caption{MDM H$\alpha$ images of the northeastern (left) and southwestern (right) outlying nebulosities around YY Hya showing the presence of emissions of overlapping shocks.
 \label{MDM_images} }
\end{center}
\end{figure*}

Higher-resolution images were also obtained with the 2.4m Hiltner telescope at the MDM
Observatory at Kitt Peak, Arizona using the Ohio State Multi-Object
Spectrograph (OSMOS; \citealt{Martini2011}) in direct imaging mode. A 4k $\times$ 4k CCD provided an effective FOV of
$18^{\prime} \times 18^{\prime}$. With $2\times2$ on-chip binning, this yielded
an image scale of $0.546^{\prime\prime}$ pixel$^{-1}$.
A series of narrow passband H$\alpha$ + [\ion{N}{ii}], [\ion{O}{III}] $\lambda$5007, and [\ion{S}{ii}] filter images were
taken of the northeastern and southwestern outlying nebulosities with exposure times of 600~s to 1200~s.
The resulting H$\alpha$ images of the NE and SW outlying nebulae are shown in Fig.~\ref{MDM_images}. The sharp filamentary appearance of the emission along with the lack of appreciable [\ion{O}{iii}] and
[\ion{S}{ii}] emissions strongly suggests these filaments are Balmer-dominated shock filaments with shock velocities $\simeq$100 km s$^{-1}$ or more depending on the density of the ambient interstellar medium.

\subsection{UV Images}

\begin{figure*}
%
%\sidecaption
%\includegraphics[width=58mm]{FUV.png}\phantom{X}\includegraphics[width=58mm]{NUV.png}
\centerline{
\includegraphics[width=90mm]{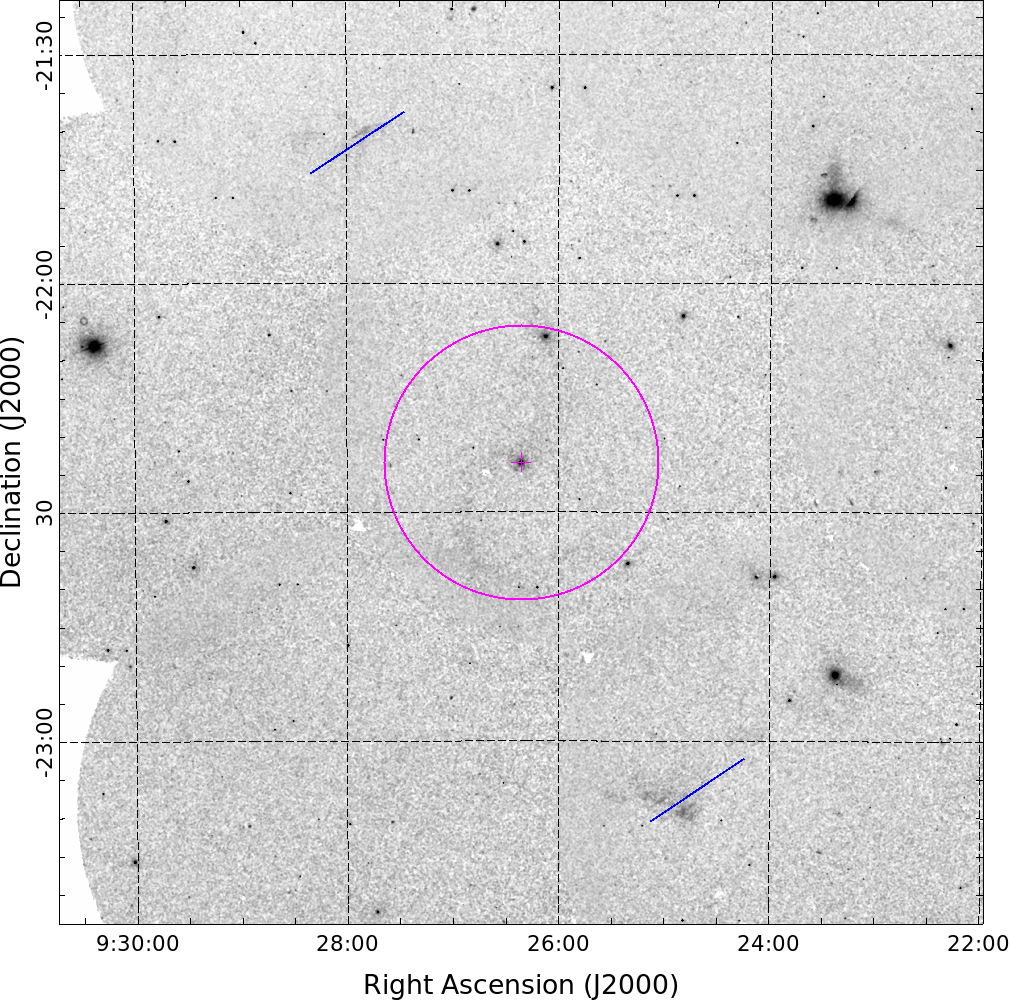}\phantom{X}\includegraphics[width=88mm]{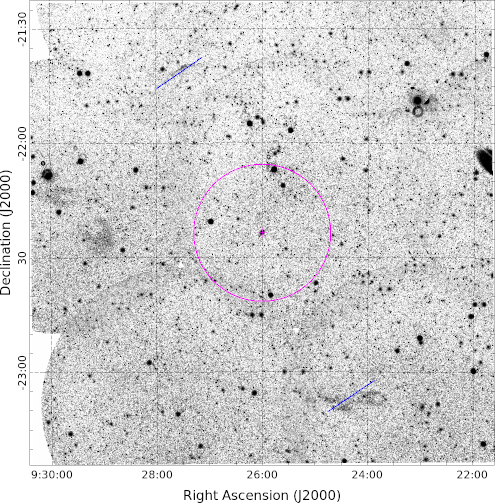}
}
\caption{The GALEX FUV (left) and GALEX NUV (right) composite mosaic image.
The 36\arcmin~ circle and the two 47\arcmin~ arcs are  centered with respect to the position of YY~Hya. While the central nebula is not clearly detected, the two vis-{\`a}-vis emission lobes are clearly visible in both bands. }
\label{fig:GALEX}
\end{figure*}

UV GALEX images \citep{GALEX_1,GALEX_2} were obtained from the online data bases. This nearly all-sky UV imaging survey used wide-band filters centered around  154.9 and 230.5\,nm, called FUV and NUV, respectively. The full-resolution images contain mainly pixels with individual photons in about 30\% of the pixels and more than 60\% of the image pixels containing no photons at all. Simply averaging these images would have artificially enhanced the intensity in the overlap regions of the individual pointings, while using a smoothing filter degrades the brighter parts of the image.  We therefore used a nearest-neighbor algorithm in the way normally used for investigations of stellar cluster density \citep{density}, which preserved the faint background without touching the brighter regions. Seven survey fields were combined that way to get a larger mosaic for the final image (Fig.\,\ref{fig:GALEX}).

Interestingly, only the far NE and SW pair of nebulae $\sim$45\arcmin\ from the center are clearly detected in the GALEX images. The average position angle of these features is 25\degr (north over east) on the sky. This is consistent with these outer emissions as being due to shocks, as suggested by the optical images. A very weak hint of parts of the main nebula in a large ''S'' shaped structure is visible.
Although GALEX images show many reflection ghost images, the detection of the lobes in the GALEX images is convincing as the features are neither in the mirroring direction along the optical axes from other bright sources, nor are they in the image of the same telescope pointing as YY~Hya. They also resemble in basic shape and location the outer nebulae seen in the optical images. The FUV/NUV $\gg$ 1 suggests similar physics of shocked gas like found in the 4~pc wide blue ring nebula around TYC~2597-735-1 \citep{BlueRing}.

\subsection{Optical and Infrared Photometry}

The light curve data from of the Catalina Real-time Transient Survey CRTS \citep{CATALINA} were downloaded from the online data base\footnote{\url{http://nesssi.cacr.caltech.edu/DataRelease/}} and converted to the solar system barycentric time frame.
Additionally, we obtained the time-series photometry from the Gaia DR2 data base \citep{GAIA_VAR}. The WISE data and a handful of data points available from the Palomar Transient Factory \citep[PTF,][]{PTF} were obtained from the NASA/IPAC Infrared Science Archive (IRSA)\footnote{\url{https://irsa.ipac.caltech.edu/frontpage/}}.

\begin{figure}[t!]
\centerline{
\includegraphics[width=88mm]{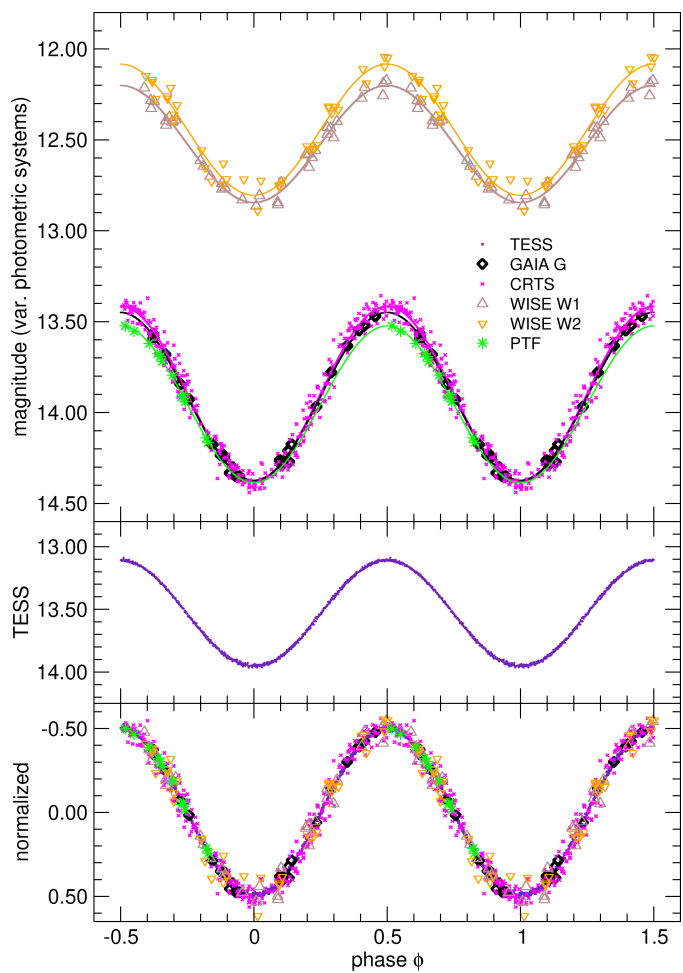}
}
\caption{The light curve taken at various facilities using the period 0\fd3347894. Upper panel: light curves of the CRTS survey, the PTF facility, WISE and Gaia have absolute calibrations. Middle panel: the TESS light curve  cannot be calibrated absolutely (see text). Lower panel: all light curves normalized to unity amplitude.}
\label{fig:lightcurve}
\end{figure}

Furthermore, we loaded the Transiting Exoplanet Survey Satellite \citep[TESS,][]{TESS} full frame images (FFI) of sector 8 (February 2020) and quick-look, early release of calibrated FFIs of sector 35 (February/March 2021) and derived the simple aperture photometry (SAP) flux light curve by using \emph{eleanor} \citep{TESS_DATA} and our in-house public tool \emph{smurfs}\footnote{\url{https://doi.org/10.5281/zenodo.3768032}}.
As the TESS pixels are very large (21{\arcsec}), the source is contaminated in the SAP aperture by the slightly brighter star Gaia DR2 5675391264067546624, which has
$G$\,=\,13\fm3697 and a color $BP-RP$\,=\,0\fm6597, very similar to that of YY Hya. Thus, while this data allows a time-series analysis, no absolute calibration of the magnitude and the amplitude is possible.
We thus used the red Gaia $RP$, with nearly the same effective wavelength of the passband (see Appendix Table~\ref{tab:photo}), to derive the calibration scale factor.

Using the visual band magnitude and color of the star in the Gaia photometry to estimate a spectral class range and the apparent magnitude, the central stellar source in the GALEX DR5 source data base \citep{GALEX_DR5} at the position of YY~Hya shows a clear UV excess.
In total, the collection of time series spans 16 years and 16\,802 periods. A complete summary of the photometry can be found in the Appendix in Table~\ref{tab:dataset}.
We searched the time series for periods using an own implementation of the phase-dispersion minimization \citep[PDM,][]{PDM} algorithm. Global solutions as well as individual blocks of 2 years were obtained to identify possible period changes, but no such changes were found down to the 0.5\,$\sigma$ level. The period, $P = 0$\fd$3347894\pm0$\fd$0000004$ ($\equiv$ to a frequency $f=2.98695\,{\rm d}^{-1}$), is constant to better than 0.1 seconds during those 16 years. Comparing the formalism from \citet{2011ApJS..194...28K}, the resulting limit of the period change implies that the system does not show strong mass transfer at the moment. The PDM is very sensitive to general period changes on those long time series. But as shown for the case of WZ~Sge by \citet{Patterson2018}, and as discussed in there for a set of other system as well, these stars often show wiggles of the observed-computed (O--C) timing of eclipses, typically on timescales of 10 to 30 years. As we do not see an eclipse in our system, we are not sensitive to such small changes here.  The ephemeris of minimum {sinusoidal optical} light {curve} is
\begin{equation}
\label{eqn:ephemeris}
% DOUBLE-CHECK BEFORE SUBMITTING.
{\rm BJD} 2458518.5082 + 0.3347894(4) E,
\end{equation}
where $E$ is the integer cycle count.
% Question - are the TESS data on the UTC or TDB time scales?  This could make a small difference in the period or epoch.
% Answer: they have an 'own' definition of barycentric time but give very accurate
% transformations for BJD including all light travel time effecs etc.
% I used their routines.
%

Even more surprising is the remarkably sinusoidal shape of the light curve. Fitting a single sine function gives for the TESS data a correlation coefficient of $R^2 \gtrapprox 0.9997$ and $rms=0\fm0073$.
The near-perfect sinusoidal light curve allowed us to normalize the amplitudes from various wavelengths to a common light curve, showing that this perfect sinusoidal behavior was valid throughout the whole data period (Fig.\,\ref{fig:lightcurve}). Moreover, after the drop from a flux ratio between maximum and minimum {of the optical light curve} of 2.5 for the two blue bands around 500\,nm to 2.2 at 600\,nm the amplitude is fairly well linear with a very small slope from 2.2 to 1.95 from approximately 610 to 4600\,{nm}. As the flux in the Rayleigh–Jeans approximation part of a spectrum depends linearly on temperature $T$, this suggests that the variability originates mostly from temperature changes similar to those found in irradiated systems \citep{EREBOS}. Near infrared (NIR) data obtained at two epochs by the Deep Near Infrared Southern Sky Survey \citep[DENIS,][]{DENIS} and one epoch from the Two Micron All Sky Survey \citep[2MASS,][]{2MASS} exist. All were obtained at phases near minimum. Using the linear relation of the amplitude with wavelength from the data with existing light curve small corrections towards an estimated minimum {light} were obtained to be used for the SED {based on ATLAS9 stellar atmospheres \citep{ATLAS9}} later. As the {NIR} observations were obtained near minimum such an estimate for observations at maximum {light} would be too much of an extrapolation. Thus we did not use the NIR photometry for the SED during maximum. The filter definitions and the calibration zero points for the entire photometric data are given in the Appendix in Table~\ref{tab:photo}.

Using the newest 3D Galactic extinction map \emph{Bayestar2019} \citep{Bayestar2019} we derive an interstellar reddening as low as $E({B-V}) = 0\fm02$ while the \emph{Stilism} map \citep{Stilism} gives a value of $E({B-V}) = 0\fm043\pm0\fm17$.
We thus adopt the mean value of 0\fm032 for our further investigations. Furthermore, the extinction curve by \cite{extinction} was adopted. This covers purely the extinction of the foreground and the environment. Possible internal extinction within the system itself cannot be detected that way.

\subsection{Optical Spectra}

We obtained spectra of YY Hya at the 2.4\,m Hiltner telescope at the MDM Observatory at Kitt Peak, Arizona, using the Ohio State Multi-Object Spectrograph (OSMOS; \citealt{Martini2011}).  A 1\farcs4 slit and a blue grism yielded a
resolution of $R$\,$\approx$\,1600 and a wavelength coverage from 3975 to 6865\,\AA.  For wavelength calibration we used
Hg, Ne, and Ar comparison lamps and fine-tuned the wavelength scale using night-sky
emission features as needed.  We observed spectrophotometric standard stars for flux
calibration.  Reductions were accomplished using an own pipeline that included elements from
astropy and IRAF/pyraf.  We also obtained a few spectra with the 1.3\,m McGraw-Hill telescope and modspec spectrograph. These had a poorer signal-to-noise ratio than the
2.4\,m data, but were consistent with the results.

Table~\ref{tab:observations} summarizes the observations. A few spectra were taken on three successive nights in 2020 December, all of them near maximum {in the optical} light {curve} (the period is so close to 1/3 day that observations on successive nights near meridian transit are at
nearly the same phase).  In 2021 February and March we obtained much more extensive
data on more widely separated nights, which sampled the entire orbit.

\begin{figure}
\centerline{
\includegraphics[width=88mm]{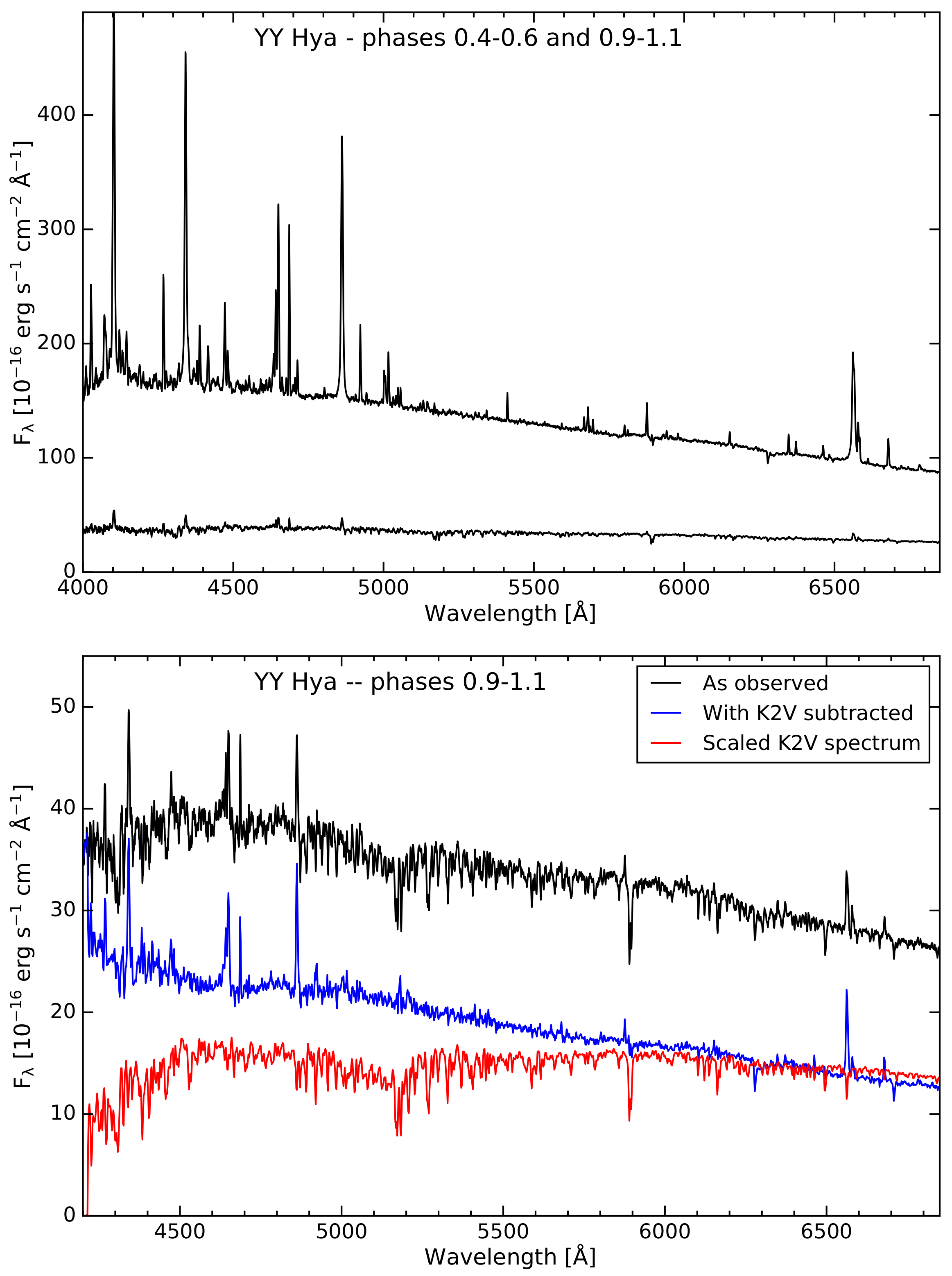}
}
\caption{{\it Upper panel:} mean OSMOS spectra from 2021 February and March; the top
trace is the average of spectra taken near maximum light, and the bottom trace near minimum light.  The vertical axis is the same for both traces. {\it Lower panel:} the
upper, black trace shows the same minimum-light spectrum as the top trace.  The red
trace is a scaled spectrum of a K2\,V star, while the blue trace results from subtracting the scaled K-type spectrum from the observed spectrum.}
\label{fig:spec_graphs}
\end{figure}

In the upper panel of Fig.~\ref{fig:spec_graphs} the averages of the flux-calibrated OSMOS spectra from near maximum and minimum {in the optical} light {curve} are plotted on the same scale. The maximum-light spectrum is dominated by a strong blue continuum and numerous emission lines of H, \ion{He}{i}, \ion{He}{ii} and weaker features that appear to
be mostly singly and doubly-ionized C and N. The latter are very narrow while the H and He lines appear to be Stark broadened.

\begin{table}% [t!]
\caption{Journal of Spectroscopic Observations obtained with the \mbox{OSMOS} spectrograph (O) at the 2.4\,m Hiltner telescope and with the \mbox{modspec} instrument (X) at the 1.3\,m McGraw-Hill telescope.}\label{tab:observations}
\begin{tabular}{c c r r r r}
\hline\hline
Start (U.T.) & \phantom{\!}\phantom{\!}\phantom{\!}Instr.\phantom{\!}\phantom{\!} & HA start\phantom{\!} & \phantom{\!}\phantom{\!}HA end~\phantom{\!}\phantom{\!} & exp. & \phantom{\!}\phantom{\!}$N_{\rm exp}$\phantom{\!} \\
             &            & \phantom{\!}(hh:mm)\phantom{\!}  & \phantom{\!}(hh:mm)\phantom{\!} & (s)~ &  \\
\hline
2020-12-15 11:10 & O & $-$00:05 & +00:26 &  600 &  3 \\
2020-12-16 10:42 & O & $-$00:30 & +00:11 & 1200 &  2 \\
2020-12-17 11:03 & O & $-$00:05 & +00:28 & 2000 &  1 \\
2021-02-07 05:47 & O & $-$01:56 & +01:55 & 1000 & 13 \\
2021-02-09 04:34 & O & $-$03:01 & +00:50 & 1000 & 13 \\
2021-03-07 03:46 & O & $-$02:07 & +02:03 & 1000 & 14 \\
2021-03-11 05:40 & X & +00:03 & +02:04 & 1800 &  4 \\
2021-03-12 05:41 & X & +00:08 & +00:38 & 1800 &  1 \\
\hline
\end{tabular}
\end{table}

This kind of mixture of wide and strong H and He lines combined with narrow emission lines of low-ionized CNO elements looks very much like that in spectra found in post common envelope pre-cataclysmic variables like \mbox{\object{EC 11575-1845}} (= \mbox{\object{TW CrV}}), \mbox{\object{V664 Cas}} \citep[the central star of the planetary nebula \mbox{\object{HFG 1}},][]{Exter2005},
\mbox{\object{HS 1857+5144}} \citep{Aungwerojwit2007,Shimansky2009},
\mbox{\object{BE UMa}} \citep{Shimansky2008},
\mbox{\object{NN Ser}} \citep[the central star of
the \mbox{\object{PN G068.1+11.0}},][]{Parsons2010, Mitrofanova2016}.
However, the emission-to-continuum line contrast is much weaker in those objects than the one we find here. Also, they all show spectra signatures of a white dwarf companion in the optical and have late M-type star companions \citep[except BE UMa, ][]{Shimansky2008}.
\mbox{\object{UU Sge}}, also belonging to that class of objects, shows predominately much higher ionization levels for the CNO elements in the spectra \citep{Wawrzyn2009}.

In Table~\ref{tab:line_list} in the Appendix a full listing including relative line strengths of the very rich emission spectrum at phase 0.5 is given. The minimum-light spectrum
is much fainter overall, with much weaker emission lines and prominent absorption
features of a late-type star.

To quantify the contribution of the late-type star, we used a set of spectra of late-type stars classified by \citet{keenan89}, observed with the modspec.  Using an
interactive program we scaled various spectra from this set and subtracted them from
the minimum-light spectra, varying the spectral type and scale factor to optimize the cancellation of the late-type absorption features. The lower panel of Fig.~\ref{fig:spec_graphs} shows the best result, obtained with a K2\,V star. Stars within $\pm 2$ subclasses of this gave acceptable cancellations.

We measured radial velocities of the late-type star using the IRAF fxcor task,
which implements cross-correlation methods described by \citet{tonry79}. As a template,
we used the average of 76 spectra of late-type IAU velocity standards that had been shifted to zero apparent velocity before the cross-correlation.  Note that in this method the wavelengths of individual features are not measured.  Regions around emission lines were masked.  The spectra near minimum light gave strong correlations.  Near maximum, in which the late-type features were barely visible, the correlations were much weaker and the velocity uncertainties correspondingly larger, but nearly all spectra showed at least some correlation.  We also measured the velocity of the H$\alpha$ emission line, using a convolution technique \citep{sy80}; the velocities of the other emission lines were similar.

\begin{figure}
\centerline{
\includegraphics[width=88mm]{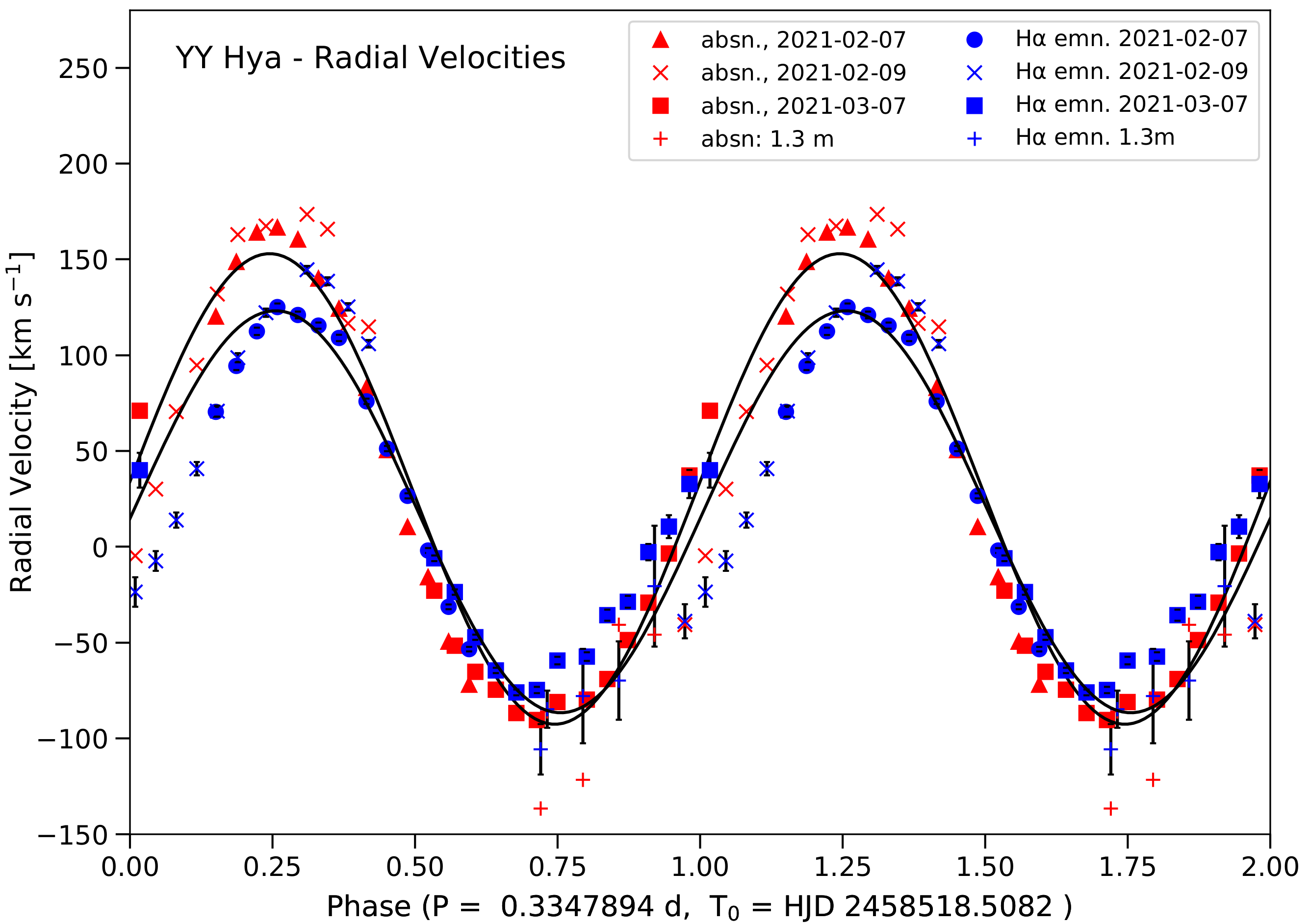}
}
\caption{Radial velocities of the absorption component and H$\alpha$ emission
line in YY Hya, as a function of the period and epoch derived from the photometry.
To preserve continuity, the data are repeated for one cycle.  The solid
lines show the best-fitting sinusoids.}
\label{fig:rvplot}
\end{figure}

Figure~\ref{fig:rvplot} shows the velocities folded with the photometric phase.
The emission lines move nearly in phase with the absorption, but with a somewhat
smaller velocity amplitude.  Table~\ref{tab:fitparams} gives the parameters of
the best fitting sinusoids of the form $v(t) = \gamma + K_{\rm em,abs}^i \sin[2 \pi(t - T_0)/P]$,
so that $T_0$ corresponds to inferior conjunction of the moving object.
The period was held fixed at the photometric value, which spans a much longer
time base; the epochs $T_0$ were allowed to vary. Even with this, both the epochs
in Table~\ref{tab:fitparams} align with the ephemeris in Eqn.~\ref{eqn:ephemeris} to within
0.01 cycles.  Equation~\ref{eqn:ephemeris} is for minimum light. The velocities
therefore indicate an irradiation effect, since the cool star's
heated face is maximally turned away from us at inferior conjunction.

\begin{table}[t!]
\caption{Radial velocities derived from the H$\alpha$ emission (em) and from numerous absorption lines (abs). The period was fixed to 0\fd3347894. 44 out of the 45 spectra from 2021 were usable.}
\label{tab:fitparams}
\begin{tabular}{l c c c c}
\hline
Data & \!$T_0$  &  $K_{\rm em,abs}^i$ & $\gamma$ & $\sigma$ \\
 & BJD &  \!\![km s$^{-1}$] & \![km s$^{-1}$] &  \!\![km s$^{-1}$] \\
 & -2400000 &  \\
\hline\hline
em & 59254.712(2) &   105(5)~ & $18(3)$  &  18 \\
abs & 59254.708(3) &   123(7)~ & $30(4)$  &  21 \\
\hline
\end{tabular}
\end{table}

\begin{figure}
\centerline{
\includegraphics[width=88mm]{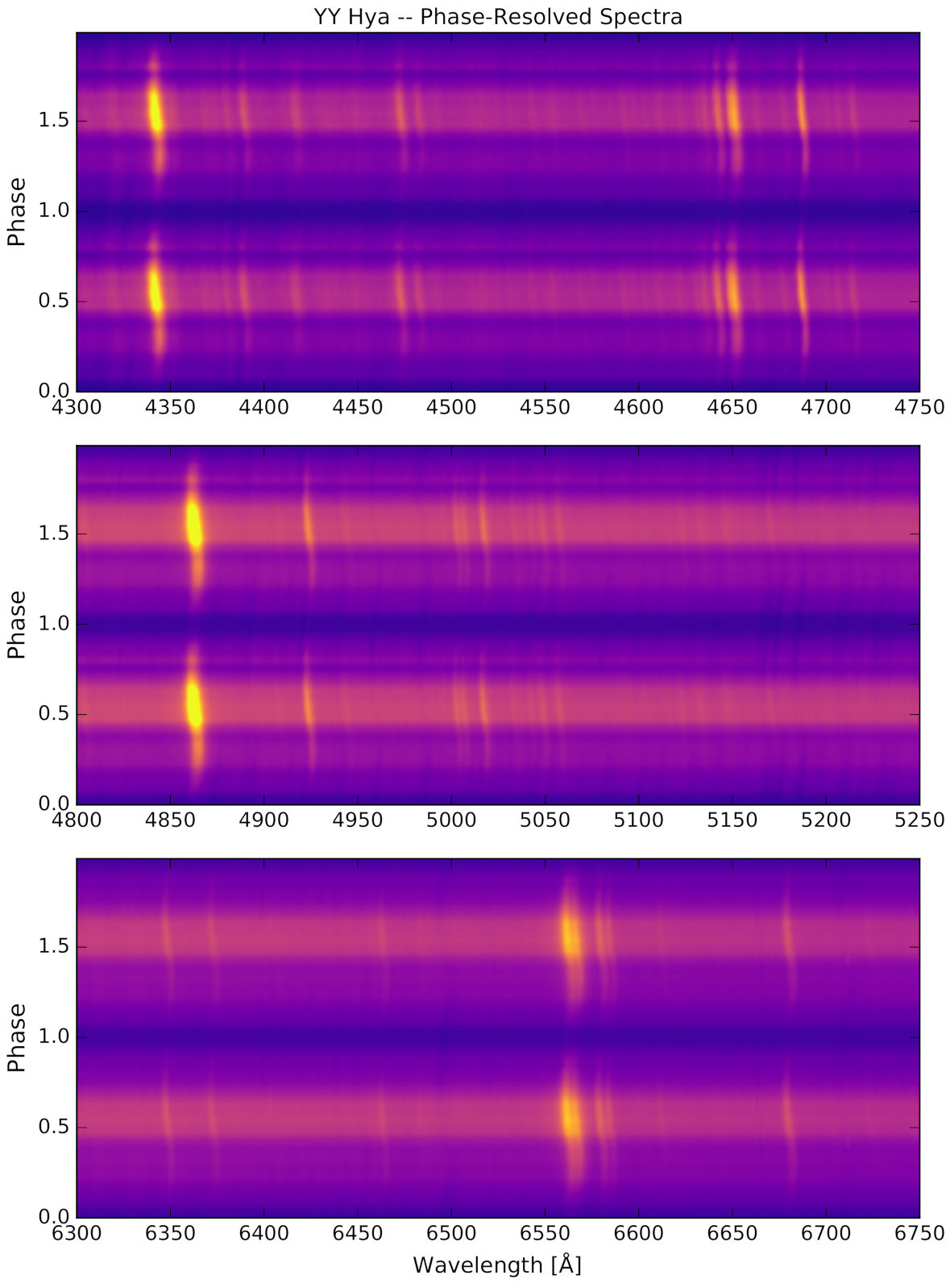}
}
\caption{Two-dimensional phase-resolved spectrum prepared as described in the text.
The spectra used here are flux-calibrated.}
\label{fig:trailplot_calib}
\end{figure}

\begin{figure}
\centerline{
\includegraphics[width=88mm]{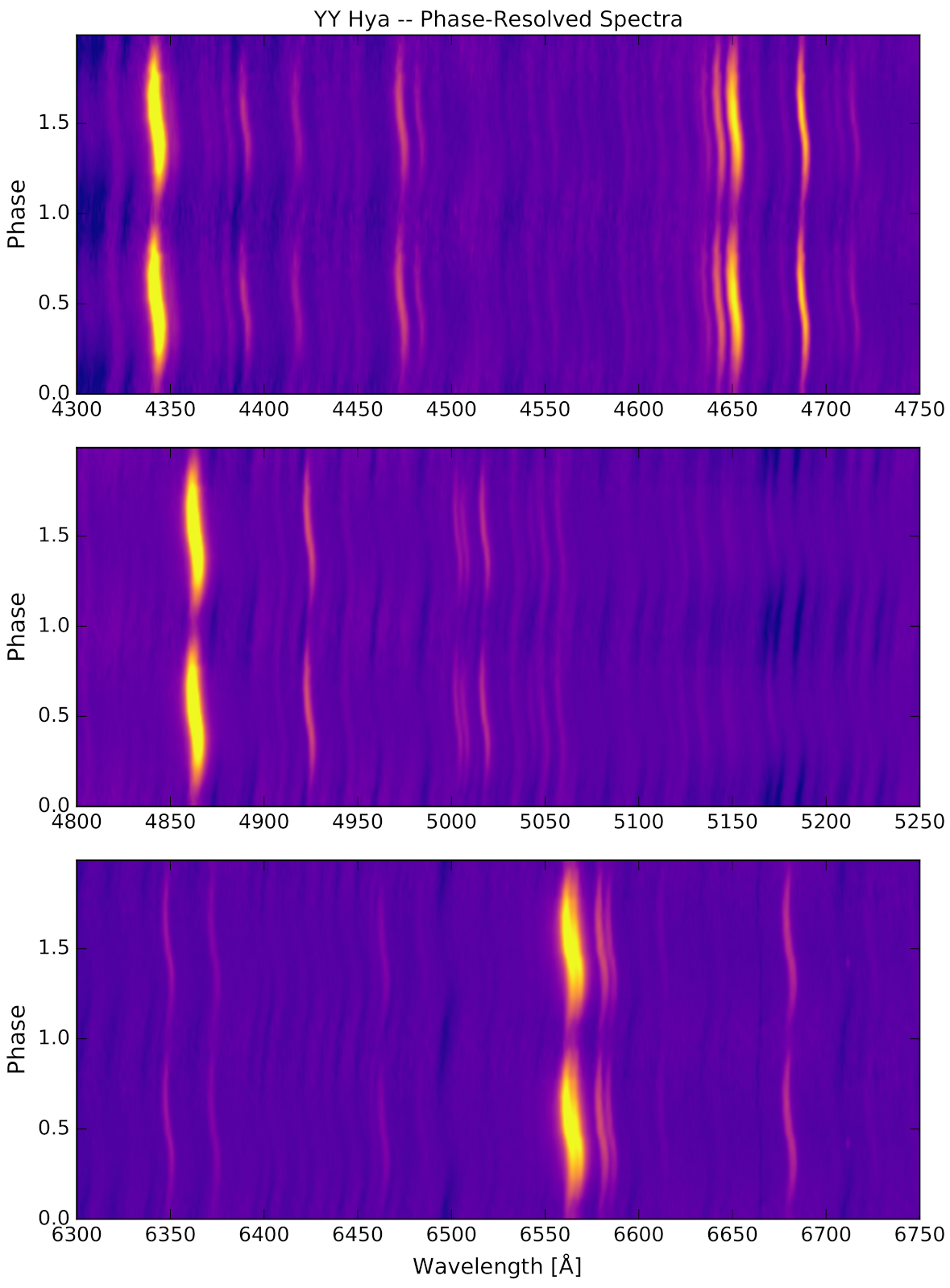}
}
\caption{Two-dimensional phase-resolved spectrum prepared as described in the text.
The spectra used here are divided by the continuum.}
\label{fig:trailplot_rect}
\end{figure}

In Figs.~\ref{fig:trailplot_calib} and \ref{fig:trailplot_rect}, we present phase-resolved spectra similar to those produced by trailing a star along a spectrograph slit. To make these, we divided the phase into 100 bins, and for each bin averaged the spectra near that phase using a narrow Gaussian window.  The spectra were then formed into a two-dimensional image, repeating once for continuity. In Fig.~\ref{fig:trailplot_calib}, we used the flux-calibrated spectra; the colormap
is set to bring out the details of the spectrum near maximum light (phase 0.5 and
1.5) and the spectrum near minimum light is invisible.  Because the light
curve is generally extremely regular, the finer-scale horizontal banding is almost
certainly an artifact of difference in the flux calibration caused by, for example,
thin cloud or variable seeing.  The emission lines dominate this view, and their velocity
variation is evident.

In Fig.~\ref{fig:trailplot_rect}, the original spectra were divided by their continua, suppressing the overall modulation but bringing up the spectra in the faint phase.  Here, the numerous K-star absorption lines can be seen, also moving approximately with the emission, as seen also in the radial velocity graphs (Fig.~\ref{fig:rvplot}).

\section{Proposed Binary Model}
\label{sec:model}

The photometric modulation and the time-resolved spectra show conclusively that YY Hya is a binary system in which a late-type star is strongly irradiated by a much hotter companion.  This, together with the UV excess, leads us to model the system as a compact binary system with a white dwarf (WD)
% or a hot sdB sub dwarf
causing the strong irradiation effects.
Other than in the objects of the BE UMa family  \citep[see e.g. ][]{Shimansky2016}, the WD is not luminous enough to contribute significantly or even to dominate the optical luminosity. The resulting WD, although providing more than 99.5\% of the flux in the GALEX FUV band, contributes less than 1\% in the optical even during minimum phase, when only the non-illuminated side of the K star is seen.
%These two types of compact companions, although originating from different evolutionary paths, occupy the same region in the photometric Hertzsprung-–Russell diagram (HRD) . The old hydrogen rich WD (DAO) tracks by \citet{DA_WD_TRACKS}, come here together with those of the extremely hot horizontal branch stars (eHB) of \citet{EHB_TRACKS}.
The irradiation modelling here follows widely the methods and discussions in \citet{V723Cas} for the post--nova V723\ Cas.

While the far side of the secondary star is near the state of the normal main-sequence star, the other side is strongly illuminated and heated. This causes the sinusoidal light curve. Such compact systems are normally in bound rotation \citep[see e.g. ][]{Schandl}. %As systems with periods and masses like we find here are partly or mostly radiative \citep{2011ApJS..194...28K}, convective mixing does drive the energy transport to the far side.
The recurrent nova CI~Aql, although even a bit more massive and with a period of $>$ 0\fd6 a significantly larger system behaves perfectly like this as well \citep{Lederle}. Similar results are found by \citet{Haefner2004} and \citet{Parsons2010} in the case of the pre-cataclysmic variable NN Ser. The latter authors write in their analysis of the light curve {\em that there is no detectable heating of the unirradiated face,
despite intercepting radiative energy from the white dwarf which exceeds its own luminosity
by over a factor of 20.0.}
Assuming now a nearly constant temperature at the far side of the cold stellar component, together with the findings in the spectroscopy (Fig.~\ref{fig:spec_graphs}), we obtained an initial guess for the system during minimum. The lack of an eclipse gives an upper boundary for the system inclination, which only marginally depends on the mass ratio $q$.
%To model the light curve in detail we used the program \emph{Binary Maker 3}
%\citep[BM3, ][]{BM3}\footnote{\url{http://www.binarymaker.com/}}.

%Using the BM3 we obtain a nearly linear rise between $i=62\degr$ to 68\degr~ for $0.3<q<1.0$ and then again %a nearly linear slow decrease to 66\degr~ for $q = 1.4$. Only extreme mass ratios cause deformations of the %Roche Lobe being large enough to cause stronger effects.
As a hot compact companion does not suffer any deformation or irradiation effects, and as it is visible always, it can be modelled as constant contribution independent from the phase. This is especially important for the GALEX FUV band, where the large secondary star does not contribute to the radiation anymore, and thus no variation is expected at all.
%The spectroscopy suggest spectral lines of a  K2V star with some additional flux (Fig. %\ref{fig:spec_graphs}).
%The latter originates from some energy transport over the illumination boundary of the
%irradiated region which is always marginally visible due to the inclination of the
%system.

Using the updated online version\footnote{\url{http://www.pas.rochester.edu/~emamajek}} of the compilation and calibration by \citet{MS_stars} we obtain then an upper limit of $M_* \le 0.8\,M_\odot$ and an radius $R_* \approx 0.75\,R_\odot$. The effective temperature then would be $T_* \le 5000$\,K. This corresponds to the spectroscopy during minimum phase shown above. However the 0.7 to 5\,$\mu$m photometric colors during the minimum phase suggest a temperature of more likely near to 4000\,K. That would correspond to a K6-7 star with $M_* \approx 0.6\,M_\odot$, $R_* \ge 0.63\,R_\odot$ and $T_* \ge 4000$\,K. We attribute the difference to the fact, that always a small fraction of the slightly heated transition zone towards the illuminated surface will show spectral lines resembling those of a slightly hotter star than that one found in the near- and mid-infrared photometry. Due to the inclination of the system here, other than in case of NN Ser, we always are contaminated by some light from the hotter illuminated side.
This is supported by the distance estimate obtained from Gaia eDR3. The sole K2V star, without additional illuminating effect, with the given magnitudes during minimum, would lie already at $> 500$\,pc. However the inclination always makes some additional flux from the illuminated side contributing to the photometry even during minimum. This effect diminishes towards the red and infrared (see also the decomposition in Fig.~\ref{fig:spec_graphs}).

The mass $M_{\rm WD}$ of the compact companion and the ratio $q$ finally is defined by the orbital period $P$, the system inclination $i$ and the velocity half amplitude $K$ from our spectroscopy by the binary mass fraction
\begin{equation}
\label{eqn:binary_mass_fraction}
f = \frac{M_{\rm WD}^3\,\sin^3\,i}{(M_{\rm WD}+M_*)^2} = \frac{P\,K^3}{2\pi\,G}
\end{equation}
where $G$ is gravitational constant, resulting in $0.4 \le M_{\rm WD} \le 1.2\,M_\odot$.

To model the light curve in detail we used the program \emph{Binary Maker 3} \citep[BM3, ][]{BM3}\footnote{\url{http://www.binarymaker.com/}}.
As mentioned already and nicely shown in the case of the evolution of the post--nova V723\ Cas over the years moving to a low quiescence state, strong mass transfers and accretion discs cause significant asymmetric structures in the light curves \citep{V723Cas}. However, as we have a perfect sinusoidal light curve, we are able to assume a system without an accretion disk here for our model.
As discussed in detail by \citet{Heber18}, irradiation efficiency is not completely understood and thus a source of systematic uncertainty. \citet{EREBOS} showed in the Eclipsing Reflection Effect Binaries from Optical Surveys (EREBOS) project with a large sample of spectroscopically investigated compact binary systems with hot subdwarf (sdB) companions that the efficiency in such systems is near to complete absorption of the UV radiation by the secondary. Model atmospheres using the PHOENIX model atmosphere code for the M-type stars of the non-mass-transferring post-common-envelope binaries
GD 245, NN Ser, AA Dor, and UU Sge show similar results \citep{Barman2004}.
Thus we adopted a unity value here. %Moreover, tests reducing the value led to no physically meaningful combination of the stellar parameters due to the high amplitude of the system.

A stochastic gradient descent method \citep{Minimum} led to various local solutions in the minimum $\chi^2$ search in the multidimensional parameter space.
Therefore we manually calculated a complete grid of light curves with inclinations varying for each mass within the above calculated range in steps of 1\degr~ and a variation of the mass  $0.4 \le M_{\rm WD} \le 1.2\,M_\odot$ in steps of 0.05\,$M_\odot$. The size $R_{\rm MS}$ of the secondary was varied from $\approx$\,80\% to exactly 100\% of the Roche Lobe $R_{\rm RL}$ in 12 steps. However, only within a few percent below the Roche Lobe perfect sinusoidal light curves were obtained. The temperature of the compact companion was varied from 50\,000 to 80\,000\,\,K. Then the luminosity of the compact star was adapted until the light curve amplitude in the TESS and the CRTS bands was recovered as observed at a phase interval of $\pm0.05$ around minimum and maximum ($\phi_{\rm MIN}=0.0$ and $\phi_{\rm MAX}=0.5$) each time. As the TESS photometric band is fairly red, this modelling does not suffer from the strong emission line regions at wavelengths below 5500\,\AA. As next step the goodness of the fit was calculated for the remaining phase regions by

$$\sum{\frac{(I_{\rm data}-I_{\rm model})^2}{\left|{I_{\rm model}}\right|}} \quad\forall (0.05<\phi<0.45) \lor (0.55<\phi<0.95). $$

\begin{figure}
%\sidecaption
\includegraphics[width=88mm]{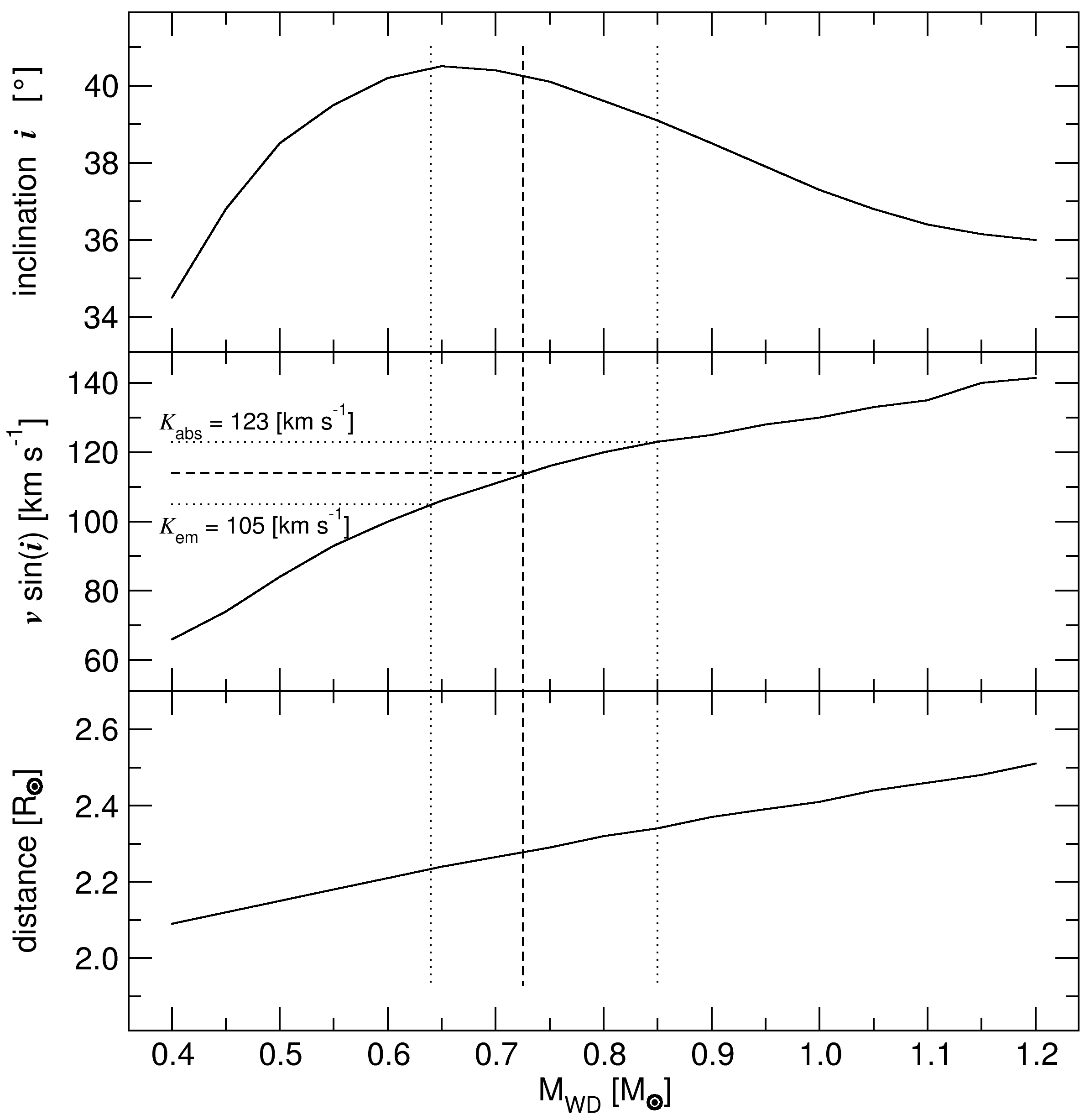}%{i_vsini_dist_B.png}
\caption{The resulting best-fit model as function of the mass of the white dwarf. The inclination $i$ of the system (top), the resulting $K$ half velocity amplitude (middle) and the orbital separation of the mass centers of the stars (bottom panel). The dashed line marks the average between the extreme cases of the emission and the absorption line velocities, while the dotted lines mark the two extreme cases (see text).}
\label{fig:M_vsini}
\end{figure}

\begin{figure}
%\sidecaption
\includegraphics[width=88mm]{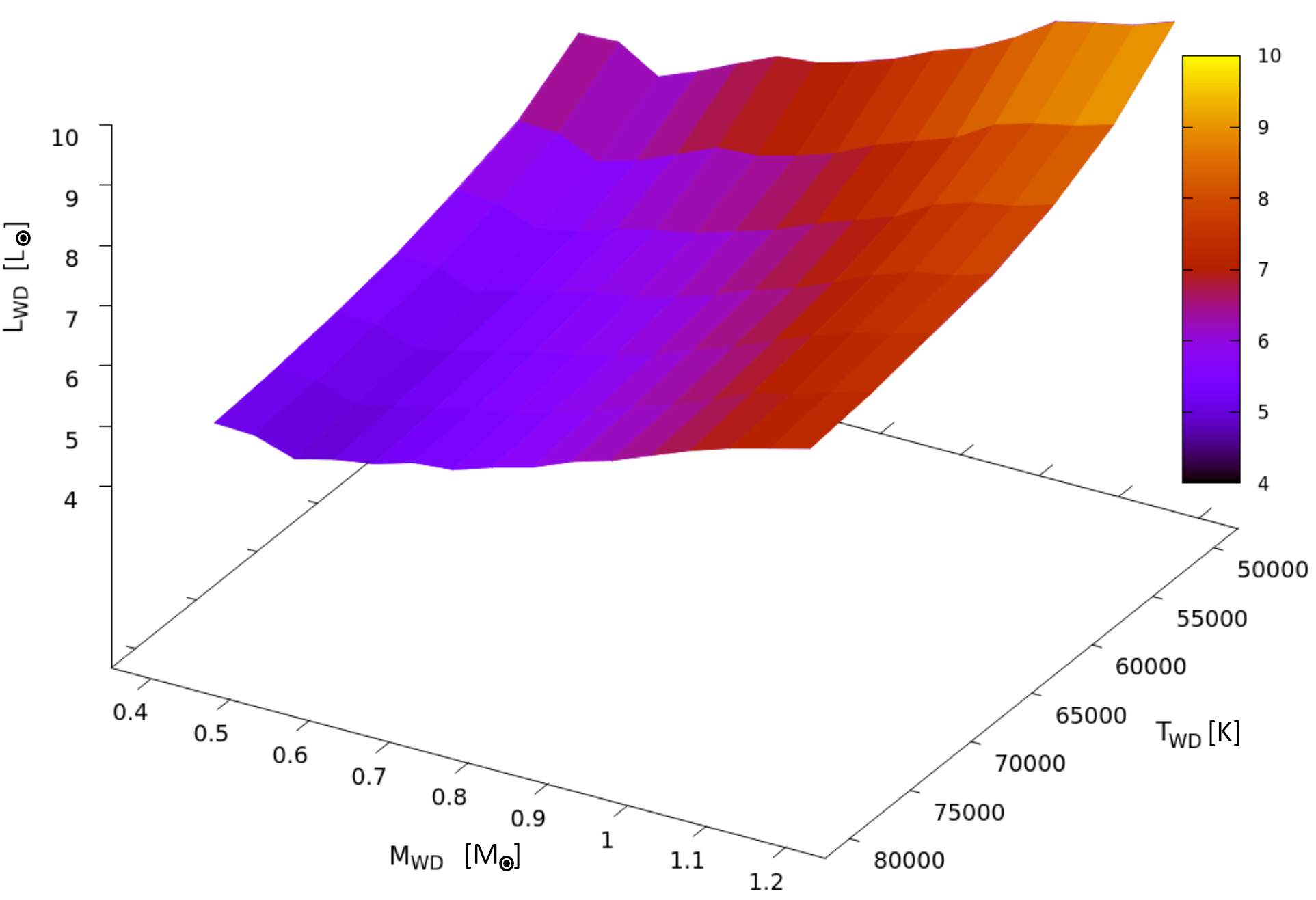}%{YY_HYA_M_T_L_plane.PNG}
\caption{The solution plane for the luminosity $L_{\rm WD}$ as function of the mass and the effective temperature of the white dwarf for the complete model grid. The luminosity is colour-coded as well for clarification.}
\label{fig:L_M_T_plane}
\end{figure}

\noindent The phase restriction for this calculation was used to avoid a dependency from the fit of the amplitude in the step before.
A small sub-sample of the residuals $(I_{\rm data}-I_{\rm model})$ of the light curves is shown in the appendix in  Fig.~\ref{fig:app_model}. Purely from photometry there is only one possible solution in inclination $i$ for each
temperature $T_{\rm WD}$ and luminosity $L_{\rm WD}$ for each of the selected $M_{\rm WD}$. Due to the nearly noise-free TESS light curve this is very sensitive to the inclination $i$. Thus around the preliminary solution the grid was refined to 0.1 degree steps. The bandwidth of solutions varies only from 36 to just above 40\degr~as function of the WD mass (Fig.~\ref{fig:M_vsini}).
The system parameters, however, give us then a radial velocity. For an illuminated star we have to expect that the emission lines originate only from the half of the star inwards to the mass center. Thus the observed velocity half-maximum $K_{\rm em}/\sin(i)$ of the emission lines will underestimate the orbital velocity $K$ from Eqn.~\ref{eqn:binary_mass_fraction}.
On the other hand the absorption lines originate from the weakly and unilluminated part, having its weighted center outside the mass center.
Thus the  $K_{\rm abs}^i/\sin(i)$ from the absorption lines will be an overestimate.
This is described also in detail for EC 11575-1845 and V664 Cas in \citet{Exter2005}. We thus safely and very conservative may use those two extreme cases $K_{\rm em}/\sin(i)$ and $K_{\rm abs}^i/\sin(i)$ as lower and upper boundaries, respectively. As indicated in Fig.~\ref{fig:M_vsini} this limits the mass range for the white dwarf to $0.64 < M_{\rm WD} < 0.85\,M_\odot$ with favorite mass of $M_{\rm WD} \approx 0.725\,M_\odot$.
The distance of the compact companion was then derived as well. The latter varies the luminosity in the mass-temperature plane (Fig.~\ref{fig:L_M_T_plane}). However, the above mass constraints limit our further steps of the investigation to a small region in the luminosity-mass-temperature plane.

\begin{figure}
\includegraphics[width=88mm]{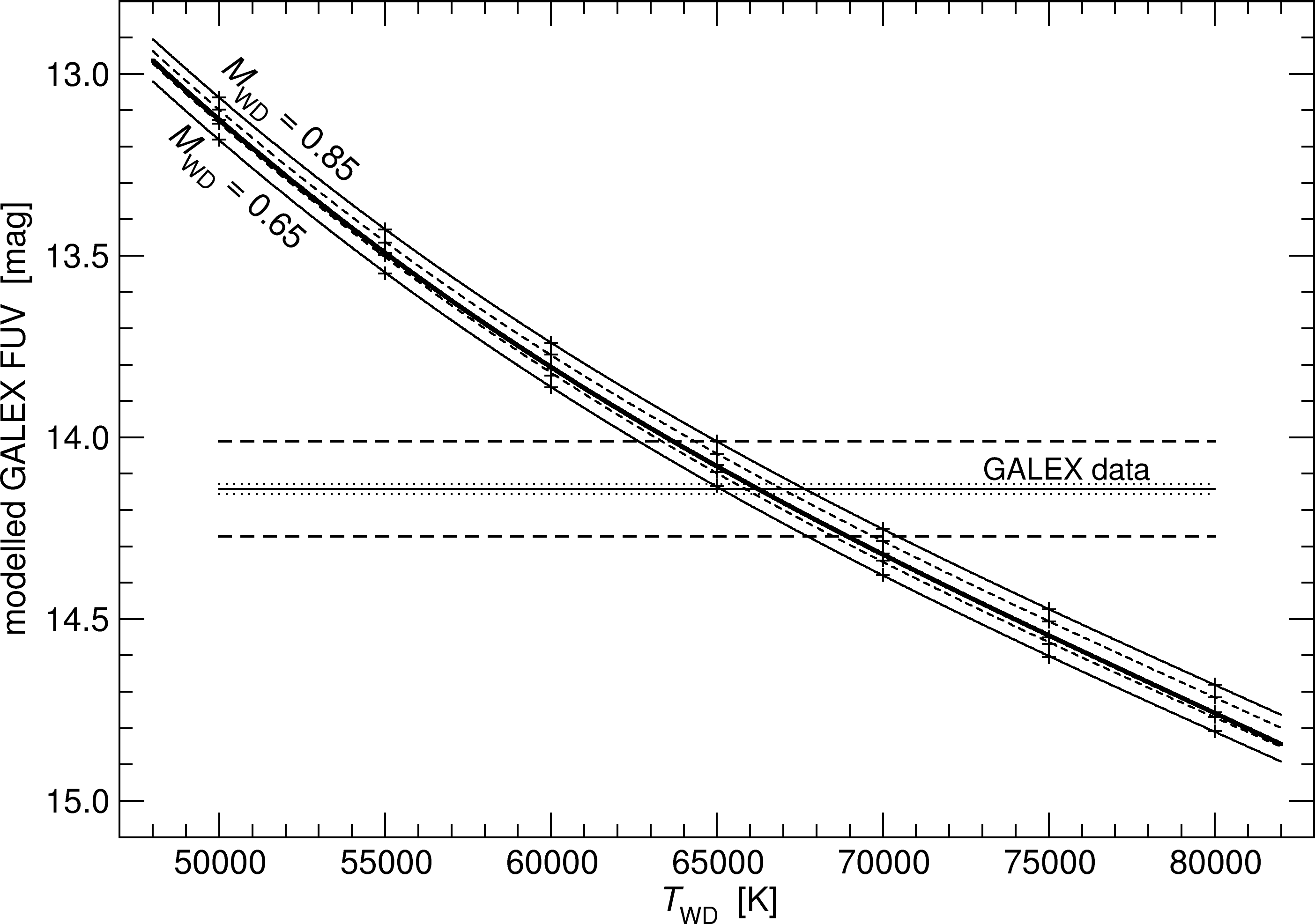}%{galex_mag_model.png}
\caption{The best model magnitude. The solution for the mass range defined from the velocities.  The GALEX measurement error (dotted lines) are very small. Conservative estimates for systematic errors are indicated by the thick dashed lines.}
\label{fig:galex_mag}
\end{figure}

%Moreover, the mass ratio $q$ defines the size of the Roche Lobe of the main sequence star %as well. As the undisturbed K star should not excess this size, this excludes at the lower %mass end of $M_{\rm WD}$ stars earlier than K6 ($M_* \approx 0.69$\,M$_\odot$, $R_* \approx 0.67$\,R$_\odot$ and $T_* \approx 4\,300$\,\,K) and thus restricts further the parameter space.
%Only at $M_{\rm WD} \ge 0.65$\,M$_\odot$ a valid solution can be found.

%%$M_{\rm WD} \approx 0.725^{+0.125}_{-0.085}$

%The resulting parameter space as function of the spectral type of the companion for different white dwarf %masses is displayed in Fig.\ %\ref{fig:inclinations}.
%This defined the boundaries of our parameter space for the modelling of the light curve.

%However, the GALEX FUV photometry restricts the $T_{\rm WD}/L_{\rm WD}$ parameter plane.

\begin{figure*}[t!]
\sidecaption
\includegraphics[width=120mm]{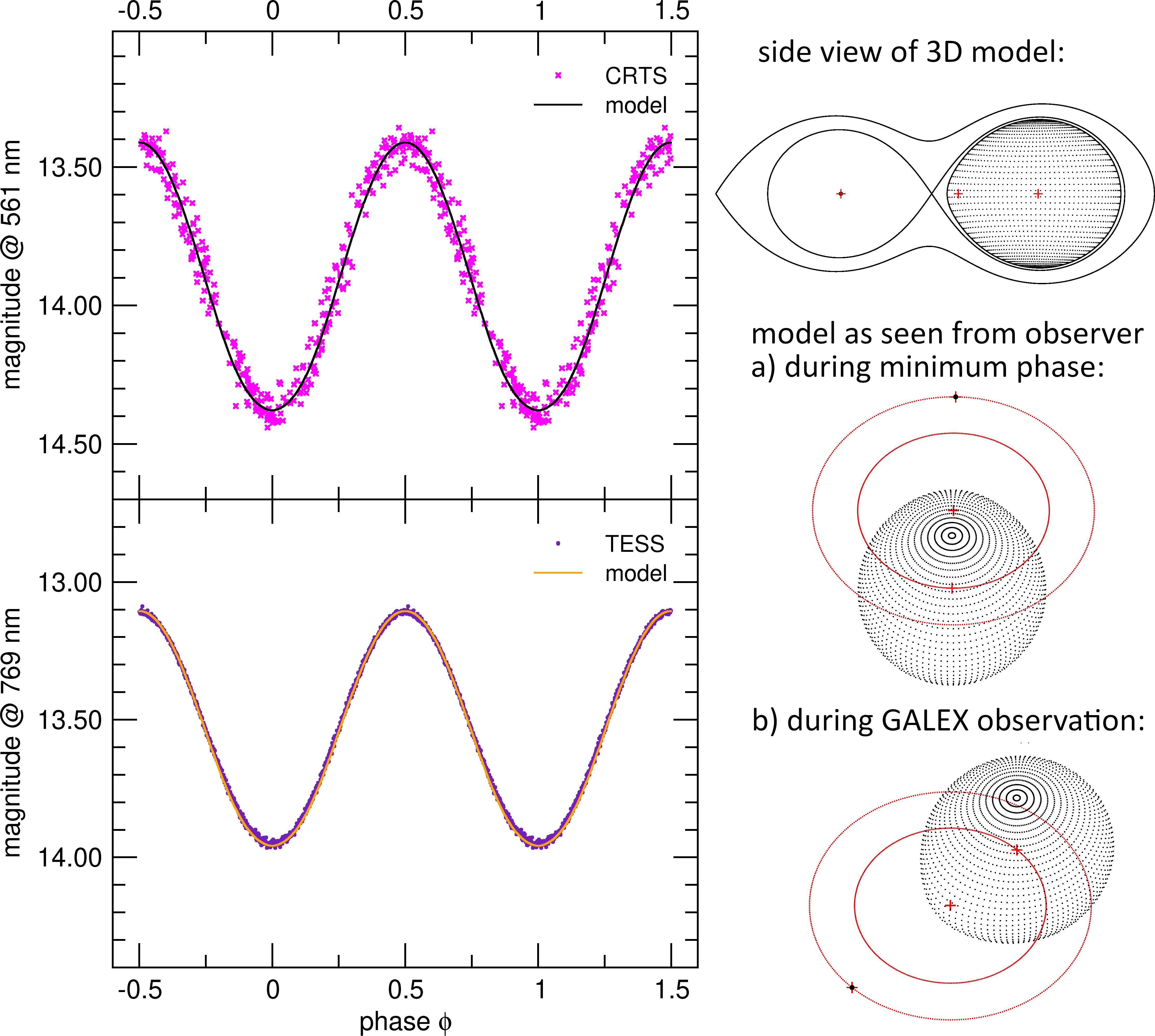}
\caption{The BM3 model. The best fit model based on the red passband TESS data and then calculated for the shorter wavelength CRTS data to constrain the temperatures of the components by the higher amplitude in the visual wavelength. The drawings (from top to bottom) show the side view including the Roche lobe limit, the model at minimum phase showing mainly the cold side of the MS star, and the phase during the GALEX observations showing nearly the whole hot side. The crosses mark the mass centers of the stars and the system barycenter. The red ellipses are the projected orbits.}
\label{fig:model}
\end{figure*}

The luminosity $L_{\rm WD}$ and the effective temperature $T_{\rm WD}$ were {\it a priori} independent free parameters, but are linked to a single degree of freedom by the GALEX FUV flux and the known distance of the system.
For this purpose all WDs with $T > 50\,000\,$K in the sample of \citet{Finley97} where GALEX FUV measurements were available were used together with the new Gaia eDR3 distances to derive a relationship between those two parameters.
For the GALEX photometry the entire model grid from 50\,000\,$<$\,$T$\,$<$\,150\,000\,K; 5.0\,$<$\,$\log(g[\mathrm{cgs}])$\,$<$\,8.0 from \citet{Rauch03} was downloaded and folded with the published GALEX FUV filter curve \citep{FILTER}. However, as the dependency on gravity $\log(g)$ is very small in our temperature range this dimension was ultimately neglected. We used for the absolute calibration the two hot WD stars 0004+330 ($T_{\rm eff}=85\,000\,$K) and 0231+050 ($T_{\rm eff}=50\,000\,$K) from the calibration sample for GALEX by \citet{GALEX_CALIB} and the WD in NN Ser due to its very accurately known stellar radius \citep{Haefner2004,Parsons2010}.
%and the GALEX data base values for the well studied stars Feige~24 ($T_{\rm eff}=62\,800\,$K) and %HD223816~B ($T_{\rm eff}=74\,000\,$K).
Although the statistical errors given for the GALEX data are very small, we used a conservative error estimate, caused from possible systematic error in the extinction correction $A_{\rm FUV}$ of 50\%.
This leads us to the final temperature range $T_{\rm WD} = 66\,500 \pm 3500\,$K (Fig.~\ref{fig:galex_mag}).
The best model parameters are summarized in Table~\ref{tab:bestmodel} and Fig.\ \ref{fig:model} shows the resulting geometry together with the light curve.

\begin{table}[t!]
\caption{Parameters of the best model.}\label{tab:bestmodel}
\begin{tabular}{l c rl l}
%    \hline
%    parameter &  &  &  & \\
    \hline
    \hline
    MS star mass & $M_{*}$ & 0.62 & $\pm0.05$ & $M_\odot$ \smallskip\\
    MS star radius & $R_{*}$ & 94\% & $^{+2\%}_{-3\%}$ & $R_{\rm RL}$ \smallskip\\
    MS star temperature & $T_{*}$ & 4.5 & $\pm0.3$ & kK \smallskip\\
    %MS star gravity\tablefootmark{a} & logg$_{*}$ & 3.35 & $^{+0.10}_{-0.08}$ & (cgs)   \smallskip\\
    MS star luminosity\tablefootmark{a} & $L_{*}$ & 0.18 & $^{+0.03}_{-0.04}$ & $L_\odot$ \medskip   \\
    WD mass & $M_{\rm WD}$ & 0.725 & $^{+0.12}_{-0.07}$ & $M_\odot$ \smallskip\\
    WD radius & $R_{\rm WD}$ & 0.019 & $^{+0.004}_{-0.003}$ & $R_\odot$ \smallskip\\
    WD temperature & $T_{\rm WD}$ & 66.5 & $\pm3.5$ & kK \smallskip\\
    %WD star gravity & logg$_{\rm WD}$ & 6.65 & $^{+0.20}_{-0.15}$ & (cgs) \smallskip \\
    WD star luminosity\tablefootmark{b} & $L_{WD}$ & 5.2 & $\pm 1.5$ & $L_\odot$ \medskip   \\
    absorption efficiency\tablefootmark{c}  & $\eta$ & 1.0 & &  \smallskip\\
    system inclination & $i$ & 40.2\degr & $\pm 0.8\degr$ & \\
     \hline
\end{tabular}
\tablefoottext{a}{\small The luminosity of the MS star was calculated from mass and\protect{\newline}\phantom{XI}\,Roche Lobe radius without taking into account other possible\protect{\newline}\phantom{XI}\,systematic effects. \protect{\newline}}
\tablefoottext{b}{\small The luminosity of the WD star is is constrained by the light curve\protect{\newline}\phantom{XI}\,amplitude. Thus error bars of the other parameters are\protect{\newline}\phantom{XI}\,not independent.\protect{\newline}}
\tablefoottext{c}{\small adopted value}
\end{table}

\section{Galactic Context}

Together with our spectroscopy, the Gaia mission data allow us to derive a full 6D dynamical investigation of the source.
We use the Gaia EDR3 parallax and proper motions, and the average of $\gamma$ derived from the emission and the absorption lines as system radial velocity with a conservative error estimate due to the scatter in Fig.~\ref{fig:rvplot} (see Table~\ref{tab:fitparams}). For the latter we thus use $v_{\rm rad}$\,=\,24$\pm$10\,km\,s$^{-1}$. We use the Galactic potential as described by \citet{AlSa91} with the code of \citet{OdBr92} to compute the Galactic orbit and kinematic parameters of YY Hya in analogy to \citet{pauli2006}.

\begin{figure*}
\sidecaption
\includegraphics[width=120mm]{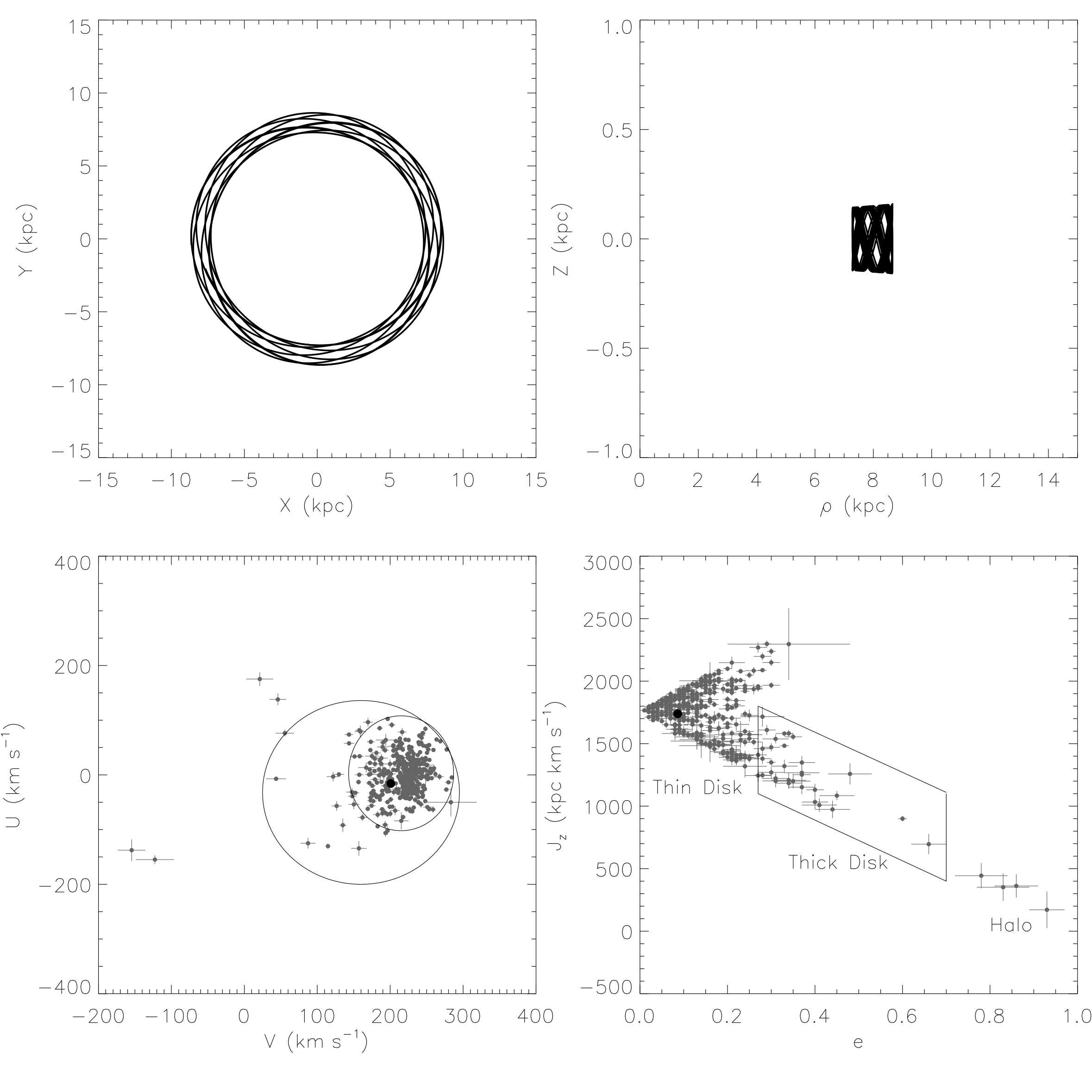}%{YYHya_kinematics.pdf}
\caption{The kinematics of YY Hya based on our spectroscopic system velocity and the Gaia parallax and proper motions. Upper panels: predicted orbit of YY Hya in the Milky Way for the next 2 Gyrs. Left: orbit projected on the Galactic plane, the Galactic center is at the origin; Right: meridional plot of the orbit. Lower panels - comparison of kinematic properties of YY Hya with the local WD sample by \citet{pauli2006}: Left: the radial velocity component $U$ versus the velocity $V$ in rotational direction, the inner circle encloses the region of thin disk objects, the outer circle the thick disk regime; Right: the angular momentum component $J_{\rm z}$ perpendicular to the plane versus the orbit eccentricity $e$.}
\label{fig:kinematic}
\end{figure*}

The orbit in the Cartesian Galactic coordinate frame $X$, $Y$, $Z$ with the Galactic center at the origin and $\rho$ parameterizing the Galactocentric distance (Fig.~\ref{fig:kinematic}) is nearly circular and bound to the Galactic disk. Moreover, the comparison with the local WD sample by  \citet{pauli2006} also show that the position of the radial versus the rotational velocity components and the angular component $J_{\rm z}$ perpendicular to the disk versus the orbital eccentricity $e$ clearly puts YY Hya into the middle of the young thin disk population. The currently reached height $Z$\,=\,155\,pc above the disk plane is in fact close to the maximum elevation reached in orbit.

Consequently, we are able to use the metal-rich WD evolution tracks from the most recent calculations by \citet{MillerBertolami2016} for the WD component of YY Hya. The WD mass of 0.725\,$M_\odot$ gives us with the initial-to-final mass relation of \citet{Cummings2018} an initial stellar mass of 3-4\,$M_\odot$. As the progenitor thus was a late B-type star \citep{mass2012,mass2018,mass2021}
with a lifetime of approximately half a Gyr or slightly below that, we conclude that the K-star companion has barely evolved off the zero-age main sequence.

%\section{Conclusion}
\section{Discussion and Conclusion}

%Although GAIA distances are often criticized for binaries.

From the data presented above, it is clear that YY~Hya is not an RR Lyr star as classified up to now in the literature but is instead a compact binary of a late K-type main-sequence star with a hot WD companion.
Due to a small separation, the system certainly went through a common envelope (CE) phase. %{\em\bf and likely had even some nova, and nova--like eruption history.}
The long term stability of the period and phase shown here, suggest that  no major mass transfer is
currently happening in this system. Moreover, the perfect symmetric and sinusoidal light curve supports that there is no accretion disk and no hot spot from an accretion stream.

Our spectroscopic results showing hardly any emission lines during minimum light phase means that its light curve is dominated by the irradiation of the compact hot WD on the secondary star. While the H and He emission lines are strongly broadened, other emission lines, mainly originating from the CNO group elements in low ionization states, are narrow. This is similar to the findings in the binaries of the BE UMa family
formed mainly by \mbox{\object{EC 11575-1845}} = \mbox{\object{TW CrV}}, \mbox{\object{V664 Cas}} = the central star of the planetary nebula \mbox{\object{HFG 1}},
\mbox{\object{HS 1857+5144}},
\mbox{\object{BE UMa}},
\mbox{\object{NN Ser}} = the central star of
the PN ETHOS 1 = \mbox{\object{PN G068.1+11.0}} \citep{Exter2005, Aungwerojwit2007,Shimansky2009, Shimansky2008, Parsons2010, Mitrofanova2016, Munday2020}.

BE UMa systems all exhibit sinusoidal light curves with no signatures of current mass transfer and active accretion. They also show spectra signatures of the white dwarf companion in the optical and have late M-type star companions \citep[except BE UMa; ][]{Shimansky2008}.
The low-luminosity companions do not outshine the WD in the optical. Thus the emission line contrast over the continuum is much smaller. The contributing WD spectra with the wide strong H and He absorptions further reduce the emission line contrast for H and He lines.
Thus all those effects we see in YY~Hya are less pronounced there. But still we denote YY~Hya as upper boundary case with higher mass companion within this class of BE UMa variables.
UU Sge, the central star of the PN A63 (PN G053.8-03.0), also belonging to that class of objects, shows predominately much higher ionization levels for the CNO elements in the spectra \citep{Wawrzyn2009}.
Older members of this physical family, with already much cooler WDs, might be \mbox{\object{MS Peg}} and \mbox{\object{LM Com}} \citep{Shimansky2003}. However, the low luminosity of the WDs in these systems result in much lower photometric amplitudes of 0\fm1 to 0\fm2 and only weak emission lines of neutral metals like \ion{Fe}{i}, \ion{Na}{i}, and \ion{Mg}{i}.

Recently \citet{CatPCE_2021} published a catalog from the literature containing more than 800 candidates of post-CE systems. Only the small fraction shown here belongs into our class of objects with periods below 1 day with no effects of the accretion of a cataclysmic variable yet. Indeed, these seem to form a class of post-CE but pre-cataclysmic variables.
 Investigations systematically searching for such close binary systems in the cores of PNe in the Gaia era will certainly expand the family \citep{Chornay2021}
 but will be limited to the young objects of the group.

We follow the suggestion by \citet{Shimansky2016}, who used the derived position along the temperature-time evolution of post-AGB white dwarfs to derive an age since the break of the AGB due to the CE phase.
With the WD mass of $0.725^{+0.12}_{-0.07}\,M_\odot$, we obtain an age since leaving the AGB of 520 to 600 kyr
from the most recent evolutionary tracks by \citet{MillerBertolami2016}.
\citet{Shimansky2016} furthermore define the relative luminosity excess $\log(\Delta L/L)$ as the difference between the expected luminosity using a WD evolutionary track and the observed one in the HRD at the post-CE age of the system.
The value of about $-$0.6 lies significantly above their relation.
However, as the reanalysis of the central star of the PN ETHOS 1 shows, moving from the very old evolutionary tracks of \citet{Bloecker1995} to the faster evolving modern post-AGB evolution change the system parameters already.
Moreover, all systems there, except the very young and hot central star of PN ETHOS 1, have much lower WD masses. Thus the initial masses of the stars, forming the compact companion later, were much lower with the secondary stars being late M stars. The exception is BE UMa. However, that system is very wide ($P_{\rm orb} = $2\fd29).

The latter difference might be the main reason why YY~Hya formed such a massive large nebula, while the young objects of the class all show nearly normal PNe. Two objects in the sample of \citet{Shimansky2016} with similar ages, namely \mbox{\object{EC 11575-1845}} and \mbox{\object{HS 1857+5144}}, however, show similar spectroscopic and photometric signatures.
Taking the size of the main nebula 4.8\,pc and assuming a CE age of 520 to 600 kyr together with typical expansions of of \mbox{5 to 20 km s$^{-1}$} for CE shells from models \citep{CEmodel1,CEmodel2}, we end up in fact with a few hundred thousand years. Moreover, \citet{CEmodel1} show that the orbital effect pronounces episodic quasi-periodic mass-loss history during CE ejection. That will give a structured nebula as we see in the case of YY Hya.

To estimate a mass we used the H$\alpha$ surface brightness using NEBULAR \citep{NEBULAR}. Assuming the shell being a aggregate of slabs with a thickness of 1\% of the radius each, and assuming the brightest fragments cover about 10\% along the line of sight in each case, we end up with a density $n_{\rm H}$ of about 2 to 7 hydrogen atoms per cm$^3$. Assuming that these fragments have a volume filling factor of  few percent we end up with a shell of about one solar mass. This supports the idea that this nebula is the result of an ejected CE. Recombination time scales of such a plasma would be 25\,000 to 100\,000 years \citep{OF06}. However, a static solution of the photoionization with the model of our WD and such a shell with CLOUDY \citep{CLOUDY} leads to photoionization fractions of about 0.5 for the hydrogen. That, and the fact, that the WD was hotter and more luminous in recent history, slows down the recombination. The CLOUDY model also predict [\ion{O}{iii}](5007\AA)/H$\alpha$ line ratios of 10$^{-4}$. Thus we failed to observe the nebula in that line. Moreover, this model would predict a [\ion{O}{ii}](3726+29\AA)/H$\alpha$ of 0.3, promising for observations.

We carried out a multi-wavelength search, using the Aladin v11 image facilities \citep{Aladin} from the radio (NVSS) over infrared (IRAS, AKARI, WISE, PLANCK), to X-rays (ROSAT) and $\gamma$-rays (FERMI). But neither the nebular center, nor the lobes appear in any of those images.

YY Hya's high Galactic latitude and the absence of large interstellar clouds in its local vicinity, the geometry of the vis-{\`a}-vis lobes plus their enormous extent, makes the explanation of ionized interstellar matter very unlikely. Instead, they are likely termination bow shocks of jets formed during the CE phase. As the FUV flux is higher in these lobes than in the main nebula, the idea that they are termination shocks of a bipolar jet is supported by
models \citep{Chamandy2018} which show that the formation of such jets is not unusual for CE evolution. However,  \citet{Chamandy2018} claim that the detailed physics near the cores still cannot be handled properly due to numerical limits on the large scale differences in the systems.
Moreover, the direction defined by the lobes point toward the Galactic plane, which is located south-west of the target. there is a slight gradient of emission visible in the PLANCK and the AKARI images. The large distance of the lobes make a gradient of the ISM already likely on the target scale. This might be the reason why the south-west lobe is so much more prominent than the north-east one.

While the wide system of BD+46\degr442 ($P$\,$>$\,140 days) certainly does not belong to this same class of CE objects, it currently shows such a jet formation  \citep{Bollen2017}. Investigations of the young objects of the EB UMa class including UU Sge \citep{Mitchell2007} and PN ETHOS 1 \citep{Miszalski2011}, as well as of PN G054.2-03.4 \citep[The Necklace;][]{Corradi2011} all show jet structures with a 3:1 size ratio compared to the main nebula. Moreover, \citet{Miszalski2011} find that all those structures are dynamically a few hundred years older than the main nebula. We note that in the case of YY~Hya such a small offset will not be significant after such a long time of $>$500\,kyr.

In summary, we find YY~Hya to be a BE UMa type post-common enevelope pre-cataclysmic variable. Within this family of objects, the  YY~Hya system extends the upper limit of mass ranges found up to now, both for the primary as
well as for the secondary star. A larger stellar shell is the reason for
the massive large emission nebula and the far distant vis-{\`a}-vis lobes.
Moreover, its high Galactic latitude ($b = +20\degr$) has helped to keep these structures
largely undisturbed for a long time. Deep and high-resolution
spectroscopy of these structures is encouraged to obtain insight into the
spatio-kinematic structure as well as more detailed information on the excitation and recombination state of such shells.

\begin{acknowledgements} We would like to thank the referee Steve Howell very much for his ideas for improving the original work. This research has made use of NASA's Astrophysics Data System Bibliographic Services (ADS), use of the SVO Filter Profile Service  supported from the Spanish MINECO through grant AYA2017-84089, NASA/IPAC Infrared Science Archive (IRSA), which is funded by the National Aeronautics and Space Administration and operated by the California Institute of Technology, and
use of the SIMBAD database \citep{SIMBAD}, operated at CDS, Strasbourg, France.
Furthermore this research made use of the Stellarium software (stellarium.org) version 0.19.1.
and of the astrometry.net project, which is partially supported by the US National Science Foundation, the US National Aeronautics and Space Administration, and the Canadian National Science and Engineering Research Council. The CSS survey is funded by the National Aeronautics and Space
Administration under Grant No. NNG05GF22G issued through the Science
Mission Directorate Near-Earth Objects Observations Program. The CRTS
survey is supported by the U.S.~National Science Foundation under
grants AST-0909182 and AST-1313422.
This publication makes use of data products from the Wide-field Infrared Survey Explorer (WISE), which is a joint project of the University of California, Los Angeles, and the Jet Propulsion Laboratory/California Institute of Technology, funded by the National Aeronautics and Space Administration. The Intermediate Palomar Transient Factory (PTF) project is a
scientific collaboration among the California Institute of Technology, Los Alamos National Laboratory, the University of
Wisconsin, Milwaukee, the Oskar Klein Center, the Weizmann
Institute of Science, the TANGO Program of the University
System of Taiwan, and the Kavli Institute for the Physics and
Mathematics of the Universe.
The MDM Observatory is operated by Dartmouth College,
Columbia University, Ohio State University, Ohio University,
and the University of Michigan.
This work presents results from the European Space Agency (ESA) space mission Gaia. Gaia data are being processed by the Gaia Data Processing and Analysis Consortium (DPAC). Funding for the DPAC is provided by national institutions, in particular the institutions participating in the Gaia MultiLateral Agreement (MLA). The Gaia mission website is \url{https://www.cosmos.esa.int/gaia}.
The Galaxy Evolution Explorer (GALEX) is a NASA Small Explorer. We gratefully acknowledge NASA’s
support for construction, operation, and science analysis for the
GALEX mission, developed in cooperation with the Centre National d’Etudes Spatiales of France and the Korean Ministry of
Science and Technology.
This paper includes data collected by the TESS mission. Funding for the TESS mission is provided by the NASA Explorer Program.
\end{acknowledgements}

%\newpage
\bibliographystyle{aa}
%\bibliography{references}

\appendix
%\onecolumn

\section{The photometry data}
\label{sec:photometry}

\begin{table}[h!]
\caption{Time span, passband definition and number of available data points for the photometric observations.}
 \label{tab:dataset}
    \begin{tabular}{lrlcr}
    \hline
    survey &  date [UTC]~~~~~~~ & ~MJD$_{\rm helio}$ & band & $N$ \\
    \hline
    \hline
         \\
    CRTS & \!\!from 2005-10-11 05:24 & \!53654.225 & $V$ & \!315 \\
     & to 2013-05-09 23:52 & \!56421.995 \\
     \\
    WISE & \!\!from 2010-05-17 06:05  & \!55333.254 & $W1$ & 37  \\
         & to 2010-11-28 22:45 &  \!55528.948 & $W2$ & 36  \\
\\
    PTF   & \!\!from 2013-03-11 04:22 & \!56362.182 & $R$ & 12  \\
 & to 2013-03-12 05:06 & \!56363.213 &   \\
\\
Gaia & \!\!from 2014-11-02 01:12  & \!56963.050 & $G$ & 27  \\
      & to 2016-05-02 19:53 & \!57510.829 & $BP$ & 27  \\
             &                  & & $RP$ & 27  \\
         \\
    TESS  & \!\!from 2019-02-05 00:21 & \!58519.015 & $R_{\rm T}$ & \!748 \\
 & to  2019-02-27 23:52 & \!58541.995 &  \\
 \\
      & \!\!from 2021-02-10 12:02 & \!59255.501 & $R_{\rm T}$ & \!\!\!2665 \\
 & to  2021-03-06 11:32 & \!59279.481 &  \\
 \\
    DENIS & at 1996-04-20 12:48& \!50193.534 & \multicolumn{2}{l}{\!$I_{\rm c}JK_{\rm s}$ }  \\
          & at 2000-01-24 18:25 & \!51567.768 & \multicolumn{2}{l}{\!$I_{\rm c}JK_{\rm s}$ }  \\
          \\
    2MASS\!\!\!~ & at 1999-02-18 14:54 & \!51227.621 & \multicolumn{2}{l}{\!$JHK_{\rm s}$ }  \\
    \\
    \hline
    \end{tabular}
\end{table}

\phantom{X}

\begin{table*}
%\begin{sidewaystable}
    \caption{Overview of the photometric bands, used effective wavelengths $\lambda_{\rm eff}$, calibration zero points $ZP$ and the interstellar extinction $A_\lambda$. Furthermore the overview of the magnitudes as used in the SED fitting (see text).
 }\label{tab:photo}
    \centering
    \begin{tabular}{lccccccccccc}
       \hline
band  & $\lambda_{eff}$ &  $ZP$ & $A_\lambda$  &  $m_\lambda^{\rm min}$ & $<m_\lambda>$  & $m_\lambda^{\rm max}$ & amplitude & phase & $m_\lambda^{\rm p}$ & $m_\lambda^{\rm min,0}$ &  \\
& [nm] & [Jy] & [mag] &  [mag] & [mag] & [mag] & [mag] & [0,1] & [mag] & \\
       \hline
       \hline
GALEX FUV         & 154.9 & ~521 & .261\,~ &        &        &        &       & 0.38 & 14.142 & 13.88 & \\
GALEX NUV         & 230.5 & ~789 & .284\,~ &        &        &        &       & 0.38 & 14.440 & 14.16 & \\
Gaia $BP$         & 502.1 & 3393 & .112\,~ & 14.577 & 14.077 & 13.576 & 1.002 &      &        & 14.46 & \\
CRTS $V$          & 561.8 & 2690 & .0975   & 14.378 & 13.897 & 13.416 & 0.963 &      &        & 14.28 & \\
Gaia $G$          & 583.6 & 2835 & .0975   & 14.374 & 13.912 & 13.449 & 0.925 &      &        & 14.28 & \\
PTF $R$           & 612.2 & 3042 & .0887   & 14.387 & 13.955 & 13.524 & 0.864 &      &        & 14.30 & \\
Gaia $RP$      & 758.9 & 2485 & .0659   & 13.958 & 13.534 & 13.110 & 0.847 &      &        & 13.89 & \\
DENIS $I_{\rm c}$ & 786.2 & 2442 & .0618   &        &        &        &       & 0.12 & 13.474 & 13.74 & \\
                  & 786.2 & 2442 & .0618   &        &        &        &       & 0.88 & 13.443 & 13.71 & \\
DENIS $J$      & 1221. & 1588 & .0293   &        &        &        &       & 0.12 & 13.115 & 13.21 & \\
                  & 1221. & 1588 & .0293   &        &        &        &       & 0.88 & 13.143 & 13.24 & \\
2MASS $J$         & 1235. & 1594 & .0288   &        &        &        &       & 0.78 & 12.867 & 13.20 & \\
2MASS $H$         & 1662. & 1024 & .0178   &        &        &        &       & 0.78 & 12.518 & 12.85 & \\
DENIS $K_{\rm s}$ & 2147. & ~667 & .0118   &        &        &        &       & 0.12 & 12.809 & 12.91 & \\
                  & 2147. & ~667 & .0118   &        &        &        &       & 0.88 & 12.660 & 12.77 & \\
2MASS $K_{\rm s}$ & 2159. & ~666 & .0117   &        &        &        &       & 0.78 & 12.392 & 12.72 & \\
WISE $W1$         & 3353. & ~310 & .0061   & 12.845 & 12.523 & 12.201 & 0.644 &      &        & 12.84 & \\
WISE $W2$         & 4603. & ~172 & .0047   & 12.805 & 12.444 & 12.084 & 0.722 &      &        & 12.80 & \\
       \hline
    \end{tabular}
%\end{sidewaystable}
\bigskip
\bigskip
\end{table*}

%\newpage
\noindent In Table\,\ref{tab:dataset} the summary for the observations for the time series used in this paper are given. Table\,\ref{tab:photo} gives an overview of the photometry. Passbands and zero points are taken from the continuously maintained online data base\footnote{\url{http://svo2.cab.inta-csic.es/theory/fps/}} of the SVO filter service \citep{FILTER}. In case of Gaia, where several filter sets are defined for DR2, we used the revised definition by \citet{GAIA_revised} from the data base. The interstellar extinction calculation is based on \citet{extinction} and the adopted reddening value of \mbox{$E({B-V})$\,=\,$0\,.\!\!^{\rm m}032$} using $R=3.1$. As the line of sight is far from the Galactic plane and does not show major dense clouds in the vicinity, these values derived for the thin interstellar matter seem to be most applicable. Due to the small value of the extinction, in fact the selection of the curve type and of $R$ is not causing a significant variation of the result. For those data sets, where a full fit of the light curve was possible, the minimum, mean and maximum from the derived sinusoidal curve fit is used. For the experiments and bands where only one or two measurements were given the phase of the measurement is given together with the magnitude $m_\lambda^{\rm p}$. Furthermore we used the wavelength--magnitude relation from the large data time series to calculate an estimate for the minimum brightness and corrected for interstellar extinction leading to $m_\lambda^{\rm min,0}$. Only for the GALEX data no such amplitude correction was derived. As discussed before we do not have to assume a significant amplitude of the the far UV component.

%\newpage
\section{The narrow-band image calibration}
\label{app:mosaic}
The image mosaic was obtained from 294 images taken at the CHILESCOPE facilities.
CHILESCOPE\footnote{\url{http://www.chilescope.com/}} is a remotely controlled commercial
observatory located in the Chilean Andes (70\degr45\arcmin53\arcsec W, 30\degr28\arcmin15\arcsec S, 1567\,m~a.s.l.) about 25\,km south of the Gemini South and LSST telescope site Cerro Pach{\'o}n.
We used the two 50cm~Newtonian telescopes built by Astro System Austria (ASA)\footnote{\url{https://www.astrosysteme.com/}}. These telescopes have a focal ratio f/3.8 and are equipped with
4\,K\,$\times$\,4\,K FLI PROLINE 16803\footnote{\url{https://www.flicamera.com/}} CCD cameras, yielding
a pixel scale of 0\farcs963 pixel$^{-1}$ and a FOV of 1\fdg096 $\times$ 1\fdg096.
Furthermore, we used for the mosaic the H$\alpha$ (FWHM 3nm) filter from Astrodon\footnote{\url{https://astrodon.com/}}.  This filter is know to suppress contamination from possible [\ion{N}{ii}] to a level below 5\% of the intensity of those lines. As the [\ion{N}{ii}]:[\ion{S}{ii}] ratio is about 10:1 in shocked nebulae like supernovae remnants and even lower for
\ion{H}{ii} regions \citep{Magrini,Barria}, and as [\ion{S}{ii}] was not detected in even the brightest regions, we are able to conclude, that we do not suffer from contamination by the nitrogen line.

Each individual exposure time was 20 minutes around 9 pointing positions with their FOV overlapping 50\% each. A total exposure time of about 100 hours was accumulated. Due to the overlap the central regions were covered with a total exposure time of 40 hours, while the outer ring with 25\% of the image width was covered only by 20 hours. However the very corners were mapped only about 11 hours. To compensate for minor variations due to weather, seeing and airmass, in the overlapping regions a common set of about 300 stars ranging from a 5$\sigma$ detection level until the overexposure at the central pixels were extracted using SExtractor v 2.19.5 (2019-07-27) \citep{SExtractor}. They were used to scale the frames  to a common flux calibration. The median of this calibration was used and the {\it rms} of these calibration factor was about 20\%. The worst case was a set of less than 10 frames achieving only about 50\% of the median flux.
due to the high overlap of up to 120 fields, no special noise suppression handling for those fields after the flux scaling was required.

The astrometry was obtained by the use of a local installation of {\sl solve-field} and {\sl wcs-resample} of the {astrometry.net}\footnote{\url{http://astrometry.net/}} suite v0.85 \citep{astrometry.net} with a Gaia eDR3 catalog subset of the surrounding 3\degr~ field. To get a proper distortion correction of this fast f/3.8 telescopes a 4$^{\rm th}$ order projection correction prior to stacking the images was required. The final positional {\it rms} was below 1/5$^{\rm th}$ of a pixel. Resampling to this grid allowed to generate a mean image now. At each position now the image count varies from 32 to 120 frames. We used the mean value of the pixels, eliminating the 3 lowest and 3 highest values.
This removed cosmics and satellite spurs but still reduces noise by building means of a large number of frames. Thus it is superior to simply using the median value.

Lacking calibrator and standard fields in the used H$_\alpha$ filter set, Gaia data were used to derive an absolute flux estimation. About 1000 stars from the central region just outside of the main nebula and from the NE and the SW corner field were matched with Gaia eDR3 stars. The red band magnitudes Gaia $RP$ and the $BP-BR$ color were used to derive the zero point $ZP$ and a linear color term

$$m_{H\alpha} = m_{\rm Gaia~RP} + k_\lambda ({BP-RP})+ ZP .$$

\begin{figure}[t!]
\sidecaption
\includegraphics[width=88mm]{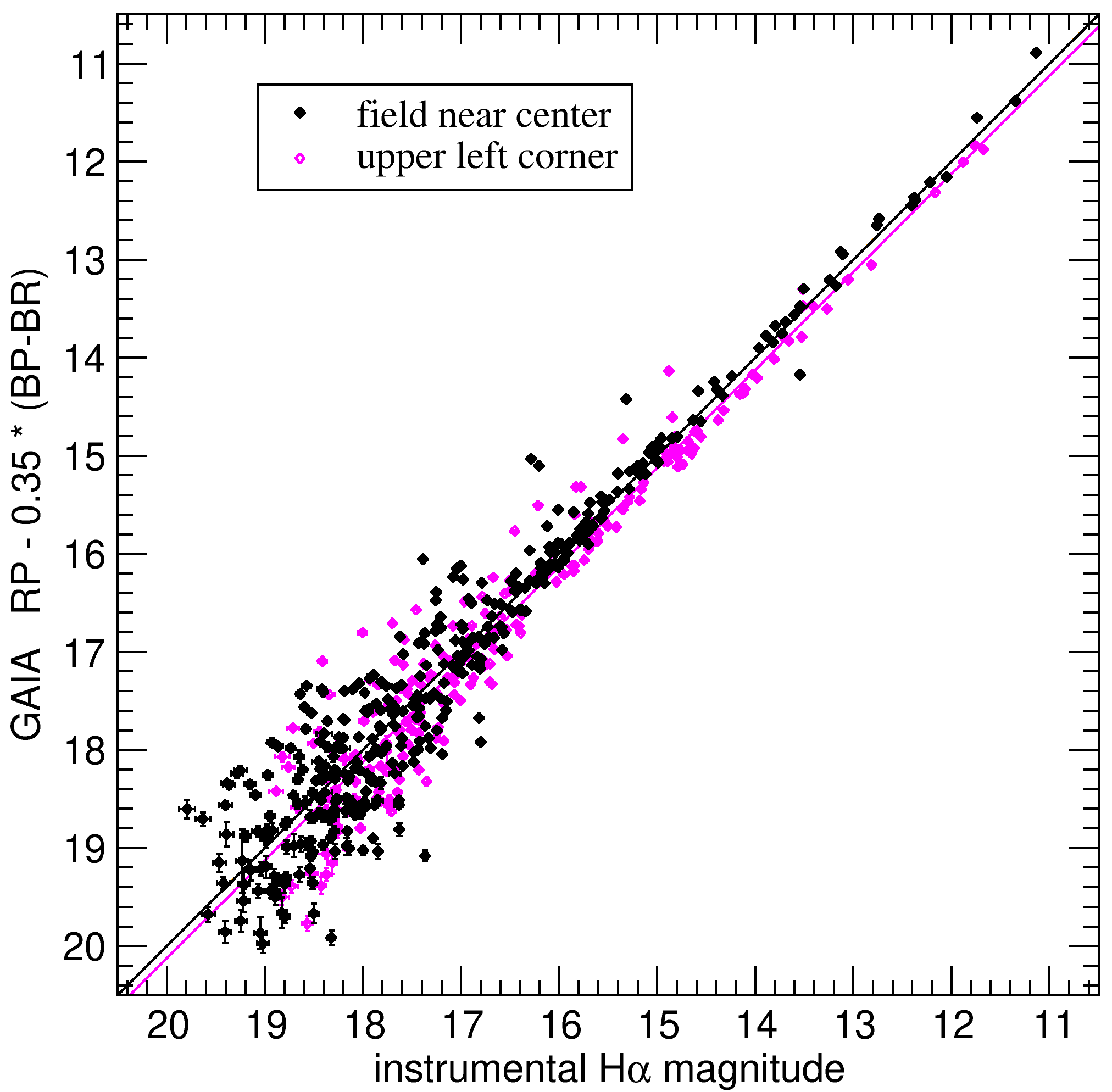}%{magnitude_calib.png}
\caption{The magnitude calibration near the field center (black) and the extreme north east corner (magenta). The Gaia catalog limited the faint end, while the CHILESCOPE H$\alpha$ goes about 1$.\!\!^{\rm m}5$ deeper.}
\label{fig:magcalib}
\end{figure}

While the $rms$ of the $ZP$ for the stars $m_{\rm Gaia~RP} < 17\fm5$ was better than 0\fm02 only minor systematic variations of the order of 0\fm08 were found between the better covered central region and the extreme field corners (Fig. \ref{fig:magcalib}). The color term $k_\lambda = 0\fm35\pm0\fm11$ was derived.
To derive a flux calibration of the emission line, the
stellar flux $f$ had to be integrated over the response curve $I$ of the 3\,nm wide filter.

$$ I_* = \int_{\lambda_1}^{\lambda_1} f(\lambda) I(\lambda) {\rm d}\lambda .$$

As the colors of the field stars show that the sample is dominated by stars later than G5 one can assume that the H$\alpha$ absorption lines do not dominate the flux variation in that region and the stars in a statistical average  and we thus can take out the $f(\lambda)$ from the integral in this small wavelength region. The response curve $I(\lambda) $ of the Astrodon filters published by the vendor was integrated numerically giving a value of 2.95\,nm. The monocromatic flux in the Vega system was obtained again from the SVO filter service \citep{FILTER} from various published 3\,nm H$\alpha$ filters from ING, ESO and TNG. They vary by less that 4\%. We adopted thus for $m_{\rm H\alpha} = 0\fm0$ the mean value of$f(\lambda) = 1.65\,\,10^{-9}$~erg\,cm$^{-2}$\,s$^{-1}$\AA$^{-1}$. Using this calibration we obtain a zero point for the surface brightness of $2.8\,\,10^{-17}$~erg\,cm$^{-2}$\,s$^{-1}$\,arcsec$^{-1}$ for the mosaic.
%The full 282\,MB image frame with 8593 $\times$ 8583 pixels can be obtained from the authors in electronic form.

\begin{table*}[ht]
\caption{Observed Emission Lines in YY Hya at Phase 0.5. The listed line strengths are relative to \ion{C}{iii} $\lambda$4650 = 100.}
\label{tab:line_list}
\centerline{
\begin{tabular}{cccllccccll}
\hline
Obs.\ $\lambda$ & Rel. & {FWHM}   & {Line}  &   {Lab $\lambda$} &
{~~~~~} &
Obs.\ $\lambda$ & Rel. & {FWHM}   & {Line}  &   {Lab $\lambda$}
 \\
$[$\AA$]$              &  Flux   & [\AA]    & {~ID}  &   ~~[\AA]  & &
[\AA]              &  Flux   & [\AA]    & {~ID}  &   ~~[\AA]  \\
\hline
\hline
 3996.4  &  3  &  3.5 & \ion{N}{ii}  &  3995.0  & & 4713.9  &  9  & 3.0  & \ion{He}{i}  & 4713.2  \\
 4010.8  &  7  &  3.9 & \ion{He}{i}  &  4009.3  & & 4803.9  &  2  & 2.9  & \ion{N}{ii}  & 4803.3   \\
 4027.6  & 41  &  4.5 & \ion{He}{i}  &  4026.2  & & 4862.0  & 178 & 7.3  & \ion{H}{i}   & 4861.4   \\
 4044.1  & 12  &  7.1 & \ion{N}{ii}  &  4041.3, 4043.53 & & 4923.0  & 26  & 3.7  & \ion{C}{i}   & 4922.7   \\
 4071.6  & 28  &  4.7 & \ion{C}{iii} &  4070.3  & & 4943.5  &  2  & 2.8  & \ion{O}{ii}   & 4943.0 \\
 4076.8  & 24  &  6.0 & \ion{C}{i}, \ion{C}{ii}  & 4073.3, 4075-76  & & 5001.9  &  7  & 2.5  & \ion{N}{ii}  & 5001.5   \\
 4103.1  & 271 &  7.7 & \ion{H}{i}   & 4101.8   & & 5006.0  & 10  & 4.2  & \ion{N}{ii}  & 5005.2, 5007.3   \\
 4121.7  & 17  &  4.2 & \ion{He}{i}  & 4120.8   & & 5016.5  & 13  & 3.1  & \ion{H}{i}  & 5115.7   \\
 4132.1  &  8  &  4.7 & \ion{N}{iii} & 4133.8, 4134.9  & &  5048.1  &  3  & 2.5  & \ion{N}{iii} & 5047.2  \\
 4145.1  & 18  &  4.6 & \ion{N}{i}, \ion{N}{ii} & 4143.4, 4145.8  & & 5056.9  &  5  & 2.9  & \ion{Si}{ii} & 5056.0 \\
 4242.1  & 20  & 15.7 & \ion{C}{i}, \ion{C}{iii} & 4223-28, 4247-52 & &  5122.5  &  2 & 2.7 & \ion{C}{ii} & 5121.8  \\
 4268.3  & 41  &  3.9 & \ion{C}{ii}  & 4267.2  & & 5132.5  &  2  & 2.7  & \ion{O}{i} & 5131.3  \\
 4277.4  &  5  &  3.5 & \ion{O}{ii}  & 4276.6  & & 5146.4  &  3  & 3.8  & \ion{C}{ii}  & 5145.2   \\
 4318.9  &  9  &  5.0 & \ion{C}{ii}  & 4317.4  & & 5169.5  &  2  & 2.2  & \ion{N}{ii}  & 5168.1  \\
 4341.7  & 227 &  7.5 & \ion{H}{i}   & 4340.5  & & 5342.9  &  2  & 2.7  & \ion{C}{iii} & 5341.5   \\
 4359.3  & 15  &  5.7 & \ion{O}{ii}  & 4357-59  & & 5412.1  &  7  & 2.8  & \ion{He}{ii} & 5411.5   \\
 4379.9  &  6  &  3.3 & \ion{C}{iii} & 4379.5  & & 5593.1  & 10  & 1.9  & \ion{N}{i}   & 5591.8  \\
 4389.0  & 28  &  4.8 & \ion{C}{iii} & 4388.0, 4390.5 & & 5667.1  &  3  & 2.4  & \ion{N}{ii}  & 5666.6   \\
 4416.6  & 22  &  5.5 & \ion{O}{ii} & 4414.9, 4417.0 & & 5679.9  &  6  & 3.0  & \ion{N}{ii}  & 5679.6   \\
 4448.8  &  2  &  2.8 & \ion{N}{ii}  &  4447.0  & & 5696.6  &  3  & 2.7 & \ion{C}{iii} & 5695.9   \\
 4472.2  & 40  &  5.0 & \ion{He}{i}  &  4471.7  & & 5801.8  &  3  & 2.9  & \ion{C}{i}   & 5800.6   \\
 4482.1  & 13  &  3.8 & \ion{Mg}{ii} &  4481.3  & & 5812.6  &  2  & 3.2  & \ion{O}{ii}   & 5810.2   \\
 4514.7  & 10  & 11.5 & \ion{N}{iii} & 4510.9, 4514.9 & & 5876.2  & 11  & 3.4  & \ion{He}{i}  & 5875.6  \\
 4532.0  &  2  & 4.9  & \ion{N}{iii} & 4530.9, 4534.6 & & 5929.8 & 3 & 6.9  & \ion{N}{ii}  & 5927.8, 5931.8   \\
 4542.5  &  4  & 4.1  & \ion{O}{iii}, \ion{N}{iii} & 4540.4, 4544.8 & & 5941.9  &  2  & 2.6  & \ion{N}{ii}  & 5941.7  \\
 4553.3  &  6  & 5.3  & \ion{N}{iii}, \ion{Si}{iii} & 4551.4, 4553-54  & & 5979.8  &  2  & 3.1  & \ion{Si}{ii} & 5979.0   \\
 4568.8  &  2  & 3.1  & \ion{Si}{iii} & 4567.8 & & 6151.8  &  4  & 3.5  & \ion{N}{ii}  & 6150.8  \\
 4591.9  &  4  & 3.2  & \ion{O}{ii} & 4591.0 & & 6347.7  &  6  & 3.4  & \ion{N}{ii}  & 6346.9   \\
 4610.2  &  5  & 4.7  & \ion{N}{ii}, \ion{N}{iii} & 4607.2, 4610.7 & & 6372.0  &  4  & 3.3  & \ion{Si}{ii} & 6371.4   \\
 4634.2  & 21  & 7.0  & \ion{N}{ii}  & 4630.5, 4634.1 & & 6462.3  & 4 & 3.7  & \ion{C}{iii}, \ion{N}{iii} & 6460.3, 6463.1 \\
 4641.9  & 35  & 3.6  & \ion{N}{iii} & 4640.6   & & 6483.1  &  1  & 6.1  & \ion{N}{ii}, \ion{N}{i} & 6482.1, 6482-84   \\
 4649.9  & 100 & 5.8  & \ion{C}{iii}  & 4647.4, 4650.3 & & 6562.7  & 65  & 8.4  & \ion{H}{i}   & 6562.85  \\
 4662.4  &  4  & 3.3  & \ion{O}{ii}  & 4661.6   & & 6578.8  & 10  & 3.5  & \ion{C}{ii}  & 6578.1  \\
 4677.1  &  4  & 2.7  & \ion{C}{i}   & 4676.7   & & 6583.5  & 6  & 3.5  & \ion{C}{ii}  & 6582.9  \\
 4686.5  & 52  & 3.4  & \ion{He}{ii} & 4685.7   & & 6611.0  & 2   & 4.0  & \ion{N}{ii}  & 6510.6  \\
 4700.0  &  2  & 2.8  & \ion{O}{ii}  & 4699.1   & & 6679.1  & 11  & 4.2  & \ion{H}{i}  & 6678.2   \\
 4705.9  &  3  & 2.7  & \ion{O}{ii}  & 4705.3   & & 6783.2  &  3  & 7.3  & \ion{C}{ii}  & 6780.3, 6783.9  \\
\hline
\end{tabular}}
\end{table*}

%\newpage
%\onecolumn
\section{Model grid}
\label{app:model}
\begin{figure*}
    \centering
    \includegraphics[width=180mm]{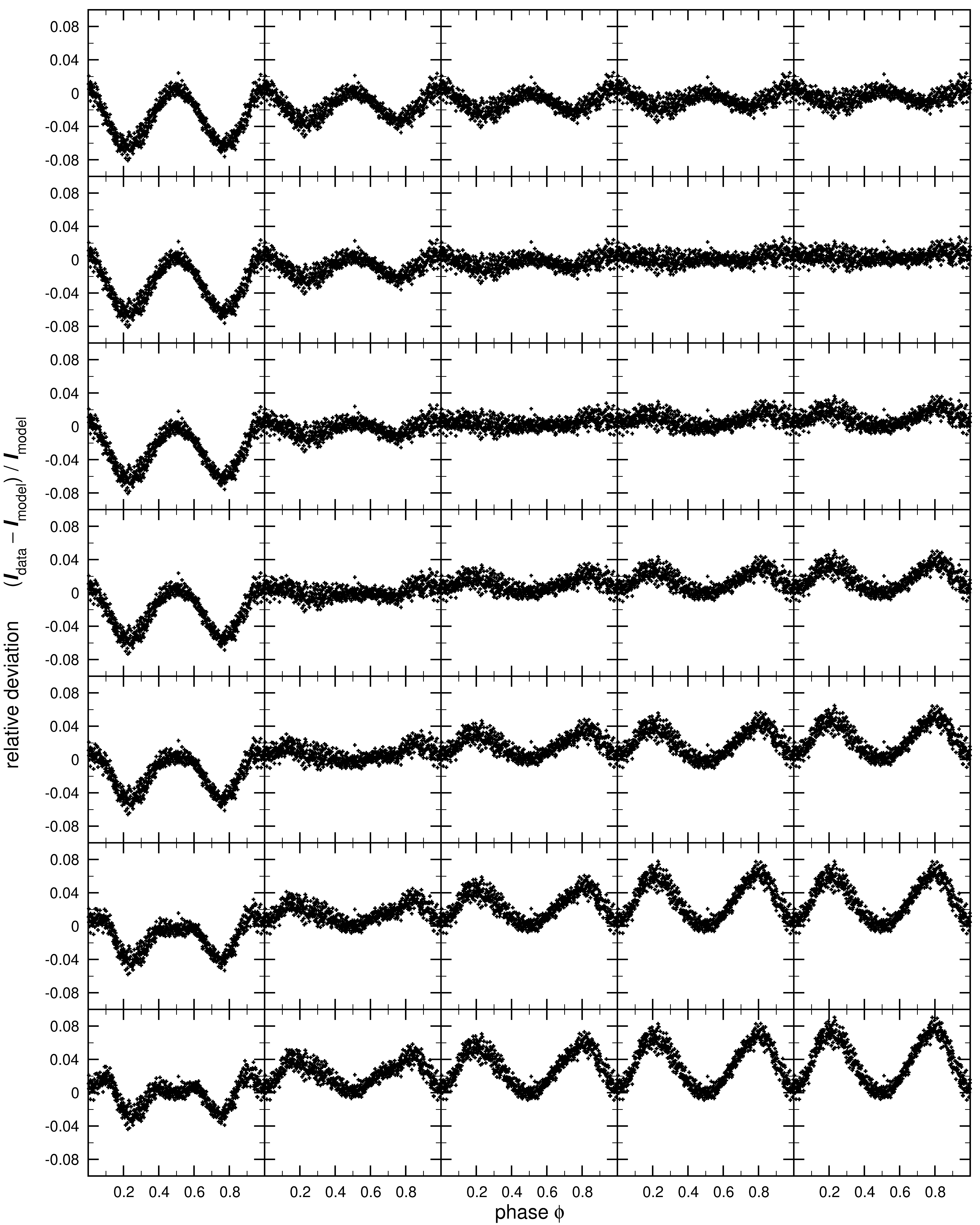}%{appendix_fig.png}
    \caption{A section of the model grid for lowest mass of the grid ($M_{\rm WD} = 0.4$\,$M_\odot$). The deviation of the TESS data intensity $I_{\rm data}$ to the model $I_{\rm model}$ as function of the phase is shown. The columns are for constant radius of the donor star from left to right 1.000, 0.955, 0.910, 0.865, 0.820 times of the Roche lobe radius $R_{\rm Roche}$. Along the rows the system inclination $i$ varies from 28\degr\ to 52\degr in steps of 4\degr. The compact companion luminosity was varied each time to fit the system amplitude (see Sect.\,\ref{sec:model}).}
    \label{fig:app_model}
\end{figure*}

Figure\,\ref{fig:app_model} shows a fraction of the whole grid of models. The effects of the most critical parameters for the modeling, namely the system inclination $i$ and the fraction of the donor star radius in units of the Roche lobe radius $R_{\rm Roche}$ are demonstrated for the solution fulfilling the best match to the amplitude for this pair of parameters.

%\relax
\newpage

\phantom{x}
\newpage

%\relax
\section{Possible Historical Link}
\begin{figure}
%\sidecaption
\centering\includegraphics[width=80mm]{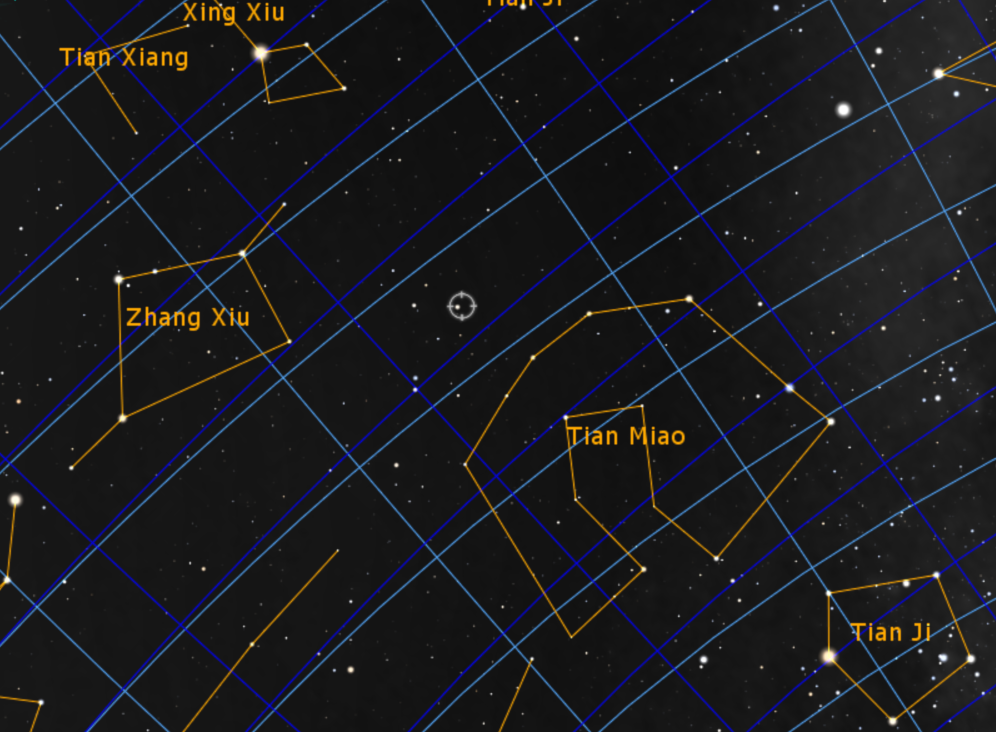}%{stellarium}
\caption{The visibility of the region of YY~Hya in the morning hours of September, 11$^{\rm th}$ 1065 from Beijing, China using the medieval Chinese constellations. {The circle marks the position of YY\ Hya}. The plot was generated by Stellarium 0.19.1}.
\label{fig:stellarium}
\end{figure}
Finally, although a bit speculative, we briefly discuss a possible link of YY~Hya and its nebula with an  apparent ``guest star'' sighting in Hydra reported in August and September 1065 AD by Korean and Chinese astronomers \citep{Hsi1957, Ho1962, Kronk1999}. Although the secondary is not Roche Lobe filling and thus is not feeding accretion yet, fallback from the very slowly ejected CE envelope may cause possibly similar accretion at slower scales on the WD.
Taking into account the rotation of the equinox,
we find YY~Hya to lie at
$\alpha=08^{\rm h}43^{\rm m}26^{\rm s}$, $\delta=-18\degr25\arcmin54\arcsec$ in 1065 AD.
Assuming an observational site in China where  the location of YY~Hya was visible above the horizon, then YY~Hya would be visible about 1.5 hours before sunrise and reached a position about 14\degr~above horizon at the end of the night. It would appear just above the constellation called the celestial temple in  the medieval  Chinese constellation map by Su Song dated 1092 \citep{StarMap}.
Using this position and constellations as input to Stellarium\footnote{\url{http://stellarium.org}}, we find that the YY~Hya is near to the constellation Tian Miao (\includegraphics[height=8pt]{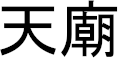} -- also called Ti{\={a}}n Mi{\`a}o, Ti{\={a}}nmi{\`a}o, Thien-miao and T'ien-Miao) containing 14 stars. {The precession-corrected position of YY Hya is indicated as well} (Fig.\,\ref{fig:stellarium}). That is what  \citep{Hsi1957} mentions for the vicinity of the 1065 event.
Furthermore, it corresponds to the maps in the investigation about ancient comets in \citet{Williams1871} and independently in the framework of the tail of the comet of 1385 in \citet{Hind1845}. They appoint this region mostly covered nowadays by Pyxis (or Pixis Nautica).

The position of YY~Hya is just on the border to this modern constellation definition\footnote{\url{https://www.iau.org/public/themes/constellations/}}. Moreover, the crude positional estimate of $\alpha=10^{\rm h}20^{\rm m}$, $\delta=-30\degr$ in B1950 ($\alpha=09^{\rm h}49^{\rm m}$, $\delta=-25\degr30\arcmin$~ in 1065) given by \citet{Hsi1957}, which is in fact
%14\degr form the position of YY~Hya and
near to the $m_V$\,=\,4\fm25 star $\alpha$~Ant, was too much south and not observable in nighttime from China at that time. We thus have not to further consider this result.
In view of these positional agreements, we propose that the link of the 1065 AD transient event with YY~Hya is at least possible.

Assuming a typical absolute visual magnitude of a Nova with $M_V \approx -8\fm0$, it should have been at $m_V \approx +0\fm2$. As there are no stars brighter than $m_V < +4\fm0$ in roughly 13\fdg5 from the position, and the nearest planet was Saturn at a distance of over 25 degrees away in late August and September 1065 AD, it would seem the 1065 AD guest star may have been fairly noticeable given this nearly empty bright star region of the sky.
However, with an angular size of 36\arcmin~ ($\equiv 4.8\,$pc @ $D_{\rm Gaia}$\,=\,456\,pc) and an age of about 1000 years, YY~Hya's main nebula would require an average expansion of slightly above 2200\,km\,s$^{-1}$ if it was created in a single event. The outer lobes lying at a distance of 11.5\,pc would lead to jets with even higher velocities, around 11\,200\,km\,s$^{-1}$. Something we do not observe. %Both sets of velocities are unusually high for a single event. Especially as Nova models suggest for low mass WD result in more massive ejecta, but velocities below 1\,000\,km\,s$^{-1}$ \citep{NovaModel}.
Also the mass estimate of the shell is by orders of magnitudes too large for a single Nova event. This, excludes that the  nebulae around YY~Hya was generated during that event about 1000 years ago. %On the other hand, these models give Nova events at very low mass loss rates of only $10^{-11}$~M$_\odot$ per year. This would fit  the model of a companion just not completely filling the Roche lobe.

%The constellations are after Su Song map of constellations dated 1092 \citep{StarMap}.

\end{document}